\documentclass[a4paper,11pt]{article}
\pdfoutput=1 

\usepackage{jcappub} 

\usepackage{longtable}
\usepackage{lscape}
\usepackage{subfigure}
\usepackage{multirow}
\usepackage{graphicx}
\usepackage{url}
\usepackage[latin1]{inputenc}

\newcommand{\lsim}{\mathrel{\mathop{\kern 0pt \rlap
  {\raise.2ex\hbox{$<$}}}
  \lower.9ex\hbox{\kern-.190em $\sim$}}}
\newcommand{\gsim}{\mathrel{\mathop{\kern 0pt \rlap
  {\raise.2ex\hbox{$>$}}}
  \lower.9ex\hbox{\kern-.190em $\sim$}}}

\newcommand{\alt}{\mathrel{\mathop{\kern 0pt \rlap
  {\raise.2ex\hbox{$<$}}}
  \lower.9ex\hbox{\kern-.190em $\sim$}}}
\newcommand{\agt}{\mathrel{\mathop{\kern 0pt \rlap
  {\raise.2ex\hbox{$>$}}}
  \lower.9ex\hbox{\kern-.190em $\sim$}}}

\newcommand{\bb}{\beta\beta}
\newcommand{\onbb}{0\nu\beta\beta}
\newcommand{\twonbb}{2\nu\beta\beta}
\newcommand{\Qbb}{$Q_{\beta\beta}$}

\title{Gaseous time projection chambers for rare event detection: Results from the T-REX project. I. Double beta decay}

\author[a]{I.~G.~Irastorza,}
\author[a,1]{F.~Aznar,\note{Present address: Centro Universitario de la Defensa, Universidad de Zaragoza. Ctra. de Huesca s/n, 50090, Zaragoza, Spain}}
\author[a]{J.~Castel,}
\author[a]{S.~Cebri\'an,}
\author[a]{T.~Dafni,}
\author[a]{J.~Gal\'an}
\author[a]{J.~A.~Garcia,}
\author[a]{J.~G.~Garza,}
\author[a, 2]{H.~G\'omez,\note{Present address: APC, Univ Paris Diderot, CNRS/IN2P3, CEA/Irfu, Obs de Paris, Sorbonne Paris Cité, France}}
\author[a]{D.~C.~Herrera}
\author[a]{F.~J.~Iguaz,}
\author[a]{G.~Luzon,}
\author[a]{H.~Mirallas,}
\author[a]{E.~Ruiz,}
\author[a, 3]{L.~Segu\'i, \note{Present address: University of Oxford, Denys Wilkinson Building, Keble Road, Oxford, OX1 3RH, UK}}
\author[a, 4]{A.~Tom\'as,\note{Present address: The Blackett Laboratory, Imperial College London, UK}}

\affiliation[a]{Grupo de F\'{\i}sica Nuclear y Astropart\'{\i}culas, Departamento de Física Teórica, \\ Universidad de Zaragoza. C/ P. Cerbuna 12 50009, Zaragoza, Spain}

\emailAdd{igor.irastorza@cern.ch}
\emailAdd{faznar@unizar.es}
\emailAdd{jfcastel@unizar.es}
\emailAdd{scebrian@unizar.es}
\emailAdd{tdafni@unizar.es}
\emailAdd{javier.galan.lacarra@cern.ch}
\emailAdd{jagarpas@unizar.es}
\emailAdd{jgraciag@unizar.es}
\emailAdd{hgomez@apc.univ-paris7.fr}
\emailAdd{dcherreramu@gmail.com}
\emailAdd{iguaz@unizar.es}
\emailAdd{luzon@unizar.es}
\emailAdd{mirallas@unizar.es}
\emailAdd{elisaruizcholiz@gmail.com}
\emailAdd{laurasgii@gmail.com}
\emailAdd{alfredo.tomas@gmail.com}

\abstract{As part of the T-REX project, a number of R\&D and prototyping activities have been carried out during the last years to explore the applicability of gaseous Time Projection Chambers (TPCs) with Micromesh Gas Structures (Micromegas) in rare event searches like double beta decay, axion research and low-mass WIMP searches. In both this and its companion paper, we compile the main results of the project and give an outlook of application prospects for this detection technique. While in the companion paper we focus on axions and WIMPs, in this paper we focus on the results regarding the measurement of the double beta decay (DBD) of $^{136}$Xe in a high pressure Xe (HPXe) TPC. Micromegas of the \textit{microbulk} type have been extensively studied in high pressure Xe and Xe mixtures. Particularly relevant are the results obtained in Xe + trimethylamine (TMA) mixtures, showing very promising results in terms of gain, stability of operation, and energy resolution at high pressures up to 10 bar. The addition of TMA at levels of $\sim$1\% reduces electron diffusion by up to a factor of 10 with respect to pure Xe, improving the quality of the topological pattern, with a positive impact on the discrimination capability. Operation with a medium size prototype of 30 cm diameter and 38 cm of drift (holding about 1 kg of Xe at 10 bar in the fiducial volume, enough to contain high energy electron tracks in the detector volume) has allowed to test the detection concept in realistic experimental conditions.  Microbulk Micromegas are able to image the DBD ionization signature with high quality while, at the same time, measuring its energy deposition with a resolution of at least a $\sim$3\% FWHM @ Q$_{\bb}$. This value was experimentally demonstrated for high-energy extended tracks at 10 bar, and is probably improvable down to the $\sim$1\% FWHM levels as extrapolated from low energy events. In addition, first results on the topological signature information (one straggling track ending in two blobs) show promising background discrimination capabilities out of reach of other experimental implementations. Moreover, microbulk Micromegas have very low levels of intrinsic radioactivity, and offer cost-effective scaling-up options. All these results demonstrate that Micromegas-read HPXe TPC remains a very competitive technique for the next generation DBD experiments.}


\begin{document}
\maketitle
\flushbottom

\section{Introduction}

Research at the frontier of physics often requires the search for phenomena of extremely low probability of occurrence. Generically referred to as ``rare event searches'', e.g., the search for dark matter WIMPs, axions or double beta decay (DBD) all share a number of experimental requirements. Being low energy events, they are buried under much higher levels of background events from environmental radiation. The task is challenging: to reduce background levels to extremely low rates and/or to ingeniously devise strategies to get distinctive signal features that help ``see the needle in the haystack''. Typical techniques associated with rare events are the use of active and passive shielding, operation in underground sites (to reduce cosmic radiation), event discrimination algorithms and a careful selection of the detector materials from the radiopurity point of view. Some particular cases pose additional requirements (like good energy resolution for DBD searches, or good energy threshold for WIMP searches). In general, the sensitivity of an experiment scales with the size of the detector, therefore good scalability prospects is a bonus for a given technique. Usually many of these requirements are contradictory and to score high in all of them is a formidable task.

With some very respectable exceptions (like for example the Gotthard TPC for DBD or the DRIFT TPC for WIMP searches, both pioneer experiments in their respective fields), gaseous Time Projection Chambers (TPCs) have not been very much used in the past for these goals because, being a gaseous target mass, it is difficult to reach large detector masses. In addition, conventional TPCs (i.e. read by multi-wire proportional counter MWPC planes) are complex detectors, electronical and mechanical-wise. On the other hand, gas TPCs could offer truly unique features of high interest for rare event detection. The topological information of the event in gas, precisely registered by an appropriately patterned readout, is a powerful tool for signal identification and background rejection. Therefore, there is a strong motivation to critically revise the limitations of TPCs and reassess their applicability to rare events, especially under the light of recent advances on TPC readouts.

Indeed, modern TPCs are progressively replacing the old MWPC readout planes with new readouts based on micropattern gas detectors (MPGD). These rely on the use of metallic strips or pads, precisely printed on plastic supports with photolithography techniques, substituting the MWPC as the amplification structure and ``sensor'' of the ionization produced in the TPC. The simplicity, robustness and mechanical precision are much higher than those of MWPCs. This, together with advances in electronics, in size, density and cost, prompted us to seriously revise the prejudice against gas TPCs for rare events.

This has been the aim of the ERC-funded T-REX project since 2009. During these years a vigorous R\&D has been carried out in generic aspects of relevance for the mentioned goal. Most of the activity has been focused on Micromegas readouts~\cite{Giomataris:1995fq}, a successful type of MPGD that -- as will be seen in the following --  shows particularly good features for low background applications. The R\&D has reviewed aspects like the radiopurity of novel MPGD fabrication techniques or that of closely-related components, the development of software and discrimination algorithms, the characterization and measurements of experimental parameters of interest, and, finally, the realization of demonstrative prototypes at the scale necessary to assess the merit of the technique.

In this and in a companion paper we present the main results obtained so far from the T-REX R\&D. Although some (but not all) of the material shown here has already been released in a number of more specific papers and theses~\cite{Cebrian:2010nw,Tomas:2009zz,Cebrian:2010ta,Andriamonje:2010zz,pacotesis,Cebrian:2012sp,ATomas_PhD,Cebrian:2013mza,Aune:2013nza,Aznar:2013jwa, lauratesis,Aune:2013pna,Herrera:2013qda,Alvarez2014a,Alvarez2014,tesisdiana,GonzalezDiaz20158,Aznar:2015iia,Iguaz:2015myh}, it is now presented in the wider context of their corresponding physics application and its potential impact in future experimental implementations is critically discussed. As will be discussed in the following, clear prospects have been opened in the fields of double beta decay, axions and low-mass WIMPs searches. For the sake of presentation, the results are organized in two papers, the first one (this one) discusses the results relevant to the search of the DBD of $^{136}$Xe, while the second one focuses on the results relevant to the search for axions and low mass WIMPs. This separation follows the different experimental features of the sought signals (MeV energies for the former, keV energies for the latter), and the corresponding experimental requirements (energy resolution for the former, energy threshold for the latter, etc.). Some of the results, however, are of common interest for all these rare event applications (e.g. the radiopurity of microbulk Micromegas readouts, described in section~\ref{sec:micromegas}).

We thus focus on DBD searches in the following. In section \ref{sec:hptpc} we introduce the primary motivations to study gas TPCs for DBD searches. In section
\ref{sec:micromegas} we describe the Micromegas readouts and the results obtained regarding their intrinsic radiopurity. In section \ref{sec:operation} we review the current experience in operating Micromegas readouts in Xenon gas, more specifically in Xe+TMA mixtures, and the main experimental parameters obtained like gain, energy resolution, and diffusion parameters. In section \ref{sec:topology} we discuss the topological information of the DBD events in a low-diffusion, highly-pixelised Micromegas TPC, while in section \ref{sec:etracks} we review the work performed in T-REX to demonstrate realistic imaging of long electron tracks in high pressure Xe. In section \ref{sec:scaling} we briefly discuss the technical scalability of this concept to very large volumes. We finish with some discussion in section~\ref{sec:discussion} and our conclusions in section~\ref{sec:conclusion}.

\section{High pressure gas TPCs to search for the DBD of $^{136}$Xe}
\label{sec:hptpc}

The search of the neutrinoless DBD of is one of the most important quests to solve the puzzle of neutrino masses. Its observation would determine the effective Majorana mass $m_{\beta\beta}$ (with some uncertainty from nuclear physics), a combination of the three neutrinos masses and mixing parameters. It would in addition confirm that the neutrino is a Majorana fermion. Combined with data from oscillation experiments, DBD experiments can provide information on the absolute mass scale of the neutrinos and their mass hierarchy (normal or inverted). Current limits on the decay half-period $T_{1/2}^{\onbb}$ are around 10$^{23}$--10$^{25}$ years depending on the isotope. Leading DBD experiments are already operating up to $\sim$100 kg of target mass, and exploring values for $m_{\beta\beta}$ of 0.2--0.4 eV. The next milestone in the field is to reach sensitivity to $m_{\beta\beta}\sim0.02$~eV, as this would allow to fully explore the inverted neutrino mass hierarchy region. For this an exposure of several ton-years of isotope is generally needed, with close to zero background counts in the energy region of interest, a formidable experimental challenge. A variety of isotopes and detection techniques are extensively being explored, but none of them has proven clear prospects to such a next generation experiment~\cite{USpanel}.

The merit of a technique to search for DBD is represented, in a simplified way, by the well-known DBD figure of merit:

\begin{equation}\label{fom_bb}
    T_{1/2}^{\onbb} \sim \frac{a\epsilon}{A}\sqrt{\frac{Mt}{b \Delta E}}
\end{equation}

\noindent where
$a$ is the isotopic abundance of the isotope of interest in the detector target mass, $\epsilon$ the detection efficiency, $A$ the atomic mass, $M$ the detector target mass, $t$ the exposure time, and $b$ the (time, mass and energy-normalized) background level, and $\Delta E$ the energy window of interest (usually defined by the energy resolution of the detector). Although the expression is approximate (among other things, it implies long enough exposure so that gaussian statistics apply to background counts), it properly shows the interplay between the different experimental parameters. Of course, beyond those parameters, aspects like cost, ease of construction and technological challenges are to be considered too. Finally, not every DBD isotope has the same potential, although isotope intercomparison is hindered by the theoretical uncertainties of the nuclear physics of the decay. In any case, in the event of a positive signal, its observation in different experiments using different isotopes will be highly advisable.

The advantages of using xenon to search for the DBD are well know. First, xenon has a natural abundance of 8.9\% in the DBD emitter $^{136}$Xe and can be enriched by centrifugation at reasonable cost. It also has a relatively high two-electron energy endpoint \cite{PhysRevLett.98.053003} ($Q_{\bb}$ = 2458 keV), and a large lifetime for the two-neutrino $\bb$ mode ($T_{1/2}^{\twonbb} \sim 2.3 \times 10^{21}$ years, as recently measured by EXO \cite{PhysRevC.89.015502} and KamLAND \cite{PhysRevC.85.045504}), which reduces the overlap of the populations of the two neutrinos and the neutrinoless modes. In addition, xenon does not have other long-lived radioactive isotopes that could contribute to the experimental background, and xenon detectors can be scaled without the presence of inner surfaces --potentially containing surface contamination (contrary to solid detectors, that are scaled by means of arraying many small modules). These features have allowed Xe experiments to take the lead in the amount of DBD isotope in operation in the last years. Examples of this are the aforementioned EXO (liquid Xe TPC) and KamLAND (Xe diluted in liquid scintillator) experiments, that have already released results at the level of $\sim$100 kg of target mass, providing leading limits to the neutrinoless DBD.

The use of a gaseous xenon TPC can bring additional advantages over a liquid detector. On one side, energy resolution is expected to improve over liquid xenon detectors \cite{Bolotnikov1997}. On the other side, detection in a gas TPC gives access --with appropriate readout granularity-- to the DBD event topology, a feature that may provide discrimination from background events. These advantages, especially the last one, may suppose a critical feature over alternative techniques in next generation DBD experiments, both to improve their sensitivity or to unambiguously identify a putative positive signal.

The first experiment exploring the high pressure Xe (HPXe) TPC concept for DBD was the Gotthard TPC \cite{Vuilleumier:1993zm} in the 90's, that operated 5 kg of Xe at 5 atm, read by a MWPC plane. Although the results were modest and the technique was considered non competitive at the time, we must acknowledge the pioneering character of the Gotthard TPC, which demonstrated important concepts like the DBD topological identification \cite{Wong:1993uq}. Almost two decades later, the concept was resuscitated within the T-REX project \cite{Cebrian:2010ta} in the form of a Micromegas-based HPXe TPC and under the light of the recent advances on TPC readouts. Similar considerations have been at the birth of the NEXT project, that since 2008 is in a prototyping stage with the final aim of building a 100 kg Xe TPC in the Laboratorio Subterráneo de Canfranc (LSC)\cite{Alvarez:2011my}. Some of the T-REX work here reviewed was done in part with the aim to provide a technological feedback at an early stage of the NEXT project. However, NEXT eventually focused on a detection concept based on the electroluminescence signal, an option that promises very good energy resolution but at the price of a reduced topological recognition capability and increased intrinsic radioactivity. This detection concept is also being followed by a Japanese group~\cite{axel_experiment}. More recently, the new PandaX-III experiment~\cite{pandaxiii_experiment} has started the project of building a 200 kg HPXe TPC based on charge readout with Micromegas. This latter project is largely based on the T-REX results here presented.

In this paper we will argue that Micromegas-based HPXe TPC is not only a competitive option for the search of DBD, but one with especially good prospects for the new generation of experiments. In order to support that claim the following issues have been addressed:

\begin{itemize}
\item Can Micromegas readouts (and the other components associated with their realistic implementation in a large-scale TPC) be fabricated with radioactivity levels low enough for the required background levels?
\item Can Micromegas readouts operate reliably in high pressure Xe, i.e. with sufficient gain, stability, homogeneity and with sufficiently good energy resolution?
\item Can the long electron tracks of the DBD events in the TPC be imaged  with sufficient 3D spatial resolution and granularity to preserve maximal discrimination power from the topological information?
\item Can a detector configuration with Micromegas readouts be realistically and cost-effectively implemented in a large scale, e.g., can large enough surface of readout be manufactured, assembled, equipped and operated?; can the associated technical requirements (e.g. gas handling, number of channels, DAQ electronics, feedthroughs, etc.) be accomplished within acceptable risk and budget?
\end{itemize}

In the following sections we describe the main results of the T-REX activities carried out to answer these questions. In general, the results obtained so far allow to answer positively these questions, and to define a competitive detector configuration that can provide close to ultimate sensitivity at the $\sim100$~kg scale. In some aspects, clear margin of improvement is still possible, and future work along specific lines may show feasibility at the ton scale too.

\section{Radiopurity of Micromegas readout planes}
\label{sec:micromegas}

The Micromegas (MICROmesh GAseous Structure) is one of the most successful MPGD, firstly conceived 20 years ago \cite{Giomataris:1995fq} and by now largely used in many areas of nuclear, particle and astroparticle physics. A relevant example of the maturity of the concept is the ongoing upgrade of the ATLAS Muon Spectrometer~\cite{Losel:2015una} for which a total of 1200 m$^2$ of Micromegas planes are being built. The Micromegas readouts are a parallel-plate configuration detector, in which a metallic micromesh is suspended over the anode plane by means
of isolator pillars, thus defining a thin amplification gap of the order of 50--150$\,\mu$m. In a TPC configuration, Micromegas planes play the role of the electron amplifying and sensing electrodes, substituting the conventional MWPC planes. Electrons drifting towards the readout, go through the micromesh holes and trigger an avalanche inside the gap, inducing detectable signals both in the anode (that is usually segmented into strips or pixels) and in the mesh. This amplification geometry presents several advantages with respect to other MPGDs, like low intrinsic gain fluctuations and a relatively low dependence of the gain on geometrical or environmental factors.

Throughout the years, the fabrication technology of the Micromegas planes has evolved significantly, and a number of fabrication techniques now exist. One of the most recent ones, the \textit{microbulk} Micromegas \cite{Andriamonje:2010zz}, has been shown to be particularly suited to rare event searches.  Microbulk Micromegas are constructed out of a (typically 50~$\mu m$-thick) kapton foil, doubly-clad with copper; on one copper plane the anode pattern (strips, pixels), while at the other one the micromesh pattern, are engraved, and the amplification gap is created by chemically removing part of the kapton layer below the mesh holes. The resulting structure is a light-weight all-in-one Micromegas plane with great uniformity of the geometrical parameters and high flexibility in pattern design. The geometrical homogeneity (in particular that of the gap) results in an improved stability and energy resolution. Indeed, typical energy resolutions achieved with microbulk Micromegas are of the order of 11(13)\%~(FWHM) at 5.9~keV in Argon-Isobutane, for unsegmented (segmented) anodes \cite{Cebrian:2010nw,Iguaz:2015myh}, numbers close to the intrinsic values for charge amplification in gas. The spatial resolution depends on the granularity of the anode, but numbers of $\sim100~\mu$m are typically achieved in $6\times 6$~cm$^2$ microbulks with 500$~\mu$m-wide strips~\cite{Aune:2009zz}.

One of the primary motivations to use microbulks for our goal has been their potentially very low levels of intrinsic radioactivity, owing to the materials they are made of (kapton and copper). Although the possibility of manufacturing radiopure version of other Micromegas types (e.g. bulk Micromegas) is still under exploration (it would bring additional advantages e.g. regarding scaling-up issues), the T-REX activity so far has been focused on the study and development of microbulks. As part of this activity, important efforts have been taken to quantify or bound the radioactivity levels of microbulk Micromegas, as well as other components of relevance that usually come along with these readouts in their implementation in realistic TPC setups (gas vessel, field cage, radiation shielding or electronic acquisition system). These efforts are mainly based on radiopurity measurements with germanium gamma-ray spectrometry performed at the LSC at a depth of 2450~meters water-equivalent using a $\sim$1~kg ultra-low background detector from the University of Zaragoza named Paquito, complemented with --for the case of some metal samples-- Glow Discharge Mass Spectrometry (GDMS) measurements\footnote{GDMS measurements carried out by Evans Analytical Group}. A detailed description and the complete results of the screening program can be found in~\cite{Cebrian:2010ta,Aznar:2013jwa,Iguaz:2015myh}. The values measured for the most relevant samples are shown in Table \ref{tab:radiopurity}, to which we refer in the following explanations.


The radiopurity of Micromegas readout planes was first analyzed in depth in \cite{Cebrian:2010ta}. A number of samples, representative of the raw materials as well as the fully manufactured readouts, were measured with Ge spectrometry. On the one hand, two samples (\#9--10 of table \ref{tab:radiopurity}) were part of fully functional Micromegas detectors: a full microbulk readout plane formerly used in the CAST experiment and a kapton Micromegas anode structure without mesh. On the other hand, two more samples (\#11--12 of table \ref{tab:radiopurity}) were just raw foils used in the fabrication of microbulk readouts, consisting of kapton metallized with copper on one or both sides. The raw materials (kapton and copper, mainly) were confirmed to be very radiopure, bounding their contamination to less than tens of $\mu$Bq/cm$^2$ for the natural U and Th chains and for $^{40}$K. The numbers for the treated foils show similar limits or values just at the limit of the sensitivity of the measurement.

Despite their importance, these bounds were still relatively modest when expressed in volumetric terms, due to the small mass of the samples. Taking advantage of their ``foil'' geometry, more sensitive measurements of some of these (and other) samples have been carried out in the BiPo-3 detector~\cite{BiPo_detector}. This detector has been developed by the SuperNEMO collaboration to measure the extremely low levels of $^{208}$Tl and  $^{214}$Bi radioactivity of the foils that hold the DBD emitter in that experiment. Currently in operation at the LSC, BiPo-3 is able to reach sensitivities down to the few $\mu$Bq/kg level by registering the delayed coincidence between electrons and alpha particles occurring in the BiPo events. This method is adequate for any sample in the form of a thin foil (below 200~$\mu$m thick, so that some alphas can escape the sample) as is the case of Micromegas readouts.

This method has been applied to a number of samples of relevance for us~\cite{BiPo_mMs} (entries \#16--20 of Table \ref{tab:radiopurity}); all of them except for \#18 are the same as the ones previously measured with Ge spectroscopy (\#10, \#12--14). For all the measured samples, only limits to the contamination in $^{208}$Tl and $^{214}$Bi can be deduced. For both cases, limits and values improve the Ge spectrometry limits by more than 2 orders of magnitude\footnote{These measurements can be translated into contamination of natural U and Th chains if secular equilibrium is assumed. BiPo-3 cannot measure $^{40}$K.}, pointing to contaminations at the level of, or below, $\sim$0.1~$\mu$Bq/cm$^2$. This is our main result so far concerning the radiopurity of Micromegas readouts, and confirms our expectations that microbulk readouts contain radioactivity levels well below typical components in very low background detectors.

A number of other samples involved in various Micromegas fabrication processes have also been measured. A kapton-epoxy foil used in the microbulk fabrication process (to join several kapton layers in more complex routing designs) has been measured in BiPo-3 (\#18 of Table \ref{tab:radiopurity}) showing similar values to the previous samples.
A sample of vacrel sheets, used in the construction of bulk Micromegas, showed good radiopurity (\#13 and \#19 of
table \ref{tab:radiopurity} respectively). This result is of interest for the development of radiopure bulk Micromegas. Also a sheet of CVD (Chemical Vapor Deposition) diamond-coated polyimide having high resistivity was screened setting upper limits for all common radioisotopes (\#14 and \#20 of table \ref{tab:radiopurity}). This result is of interest for the development of resistive microbulk Micromegas.
However, large activities were measured for a sample of ceramic material intended to be used in ``piggyback'' resistive Micromegas (\#15 of table \ref{tab:radiopurity}), a novel implementation proposed in~\cite{piggyback}, having a substrate mainly made of alumina.


In addition to the above results regarding materials directly involved in the constitution of Micromegas readout planes, an extensive measurement campaign is being performed addressing other components or materials that are typically associated with the implementation of Micromegas readouts in actual TPCs. Detailed results can be found in~\cite{Aznar:2013jwa,Iguaz:2015myh}. For the sake of completeness, we summarize below the most relevant results.

\begin{itemize}

\item Samples from different suppliers of lead, used for shielding, and copper, used for mechanical and electric components (like Micromegas plates, cathodes, HV feedthroughs or field cage rings), have been analyzed. Very stringent upper limits were obtained by GDMS for Oxygen Free Electronic (OFE, C10100) copper from Luvata company (\#1 of table \ref{tab:radiopurity}).

\item Material and components to be used inside the gas vessel, mainly related to the field cage, have been screened and therefore adequate teflon, resistors or adhesives have been selected. The monolayer Printed Circuit Board (PCB) made of kapton and copper for the field cage, supplied by LabCircuits, was found to have good radiopurity (\#2 of table \ref{tab:radiopurity}). For the epoxy resin Hysol RE2039 from Henkel no contaminant could be quantified (\#3 of table \ref{tab:radiopurity}) and it is being used for gluing. Surface Mount Device (SMD) resistors supplied by Finechem showed lower activity than other equivalent units (\#4 of table \ref{tab:radiopurity}) and are components of choice for field cage construction.

\item Electronic connectors made of Liquid Crystal Polymer (LCP) have shown unacceptable activities of at least several mBq/pc for isotopes in $^{232}$Th and the lower part of $^{238}$U chains and for $^{40}$K. For connectors made of silicone (Fujipoly Gold 8000 connectors type C) a lower activity of $^{226}$Ra and especially of $^{232}$Th was found (\#5 of table \ref{tab:radiopurity}) and their use is foreseen. Very radiopure, flexible, flat cables made of kapton and copper have been developed in collaboration with Somacis, performing a careful selection of the materials included and avoiding glass fiber-reinforced materials at base plates. After the screening of several cable designs, the good results obtained for the final one (\#6 of table \ref{tab:radiopurity}) allow to envisage the use of these materials also at Micromegas production. Several kinds of high voltage or signal cables have been analyzed too. A sample of a cable from Druflon made of Silver Plated Copper wires with a teflon jacket was screened, showing much better radiopurity than typical RG58 coaxial cables (\#7 of table \ref{tab:radiopurity}). Different materials can be taken into consideration for PCBs and samples of FR4, ceramic-filled PTFE composite and cuflon were screened. The first ones presented very high activities for the natural chains and $^{40}$K, precluding its use. Good radiopurity was found for cuflon from Crane Polyflon; however, its application for micromegas has been disregarded due to the difficulty to fix the mesh and also because bonding films to prepare multilayer PCBs have been shown to have unacceptable activity \cite{1748-0221-10-05-P05006}. Following the measurements \#2 and \#6 of table \ref{tab:radiopurity}, kapton-copper boards seem to be the best option.


\end{itemize}

\footnotesize
\begin{landscape}
\begin{longtable}{p{0.2cm}p{4.1cm}p{1.2cm}p{1.2cm}p{1.4cm}p{1.5cm}p{1.5cm}p{1.5cm}p{1.3cm}p{1.7cm}p{0.9cm}p{0.9cm}}
 \hline
\textbf{\#} & \textbf{Material, Supplier} &   \textbf{Method} &\textbf{Unit} & \textbf{$^{238}$U} & \textbf{$^{226}$Ra} & \textbf{$^{232}$Th} &\textbf{$^{228}$Th} & \textbf{$^{235}$U}& \textbf{$^{40}$K}  & \textbf{$^{60}$Co}& \textbf{$^{137}$Cs}\\
 \hline
\endfirsthead
(Continuation)\\
\hline
\textbf{\#} & \textbf{Material, Supplier} &  \textbf{Method} & \textbf{Unit} & \textbf{$^{238}$U} & \textbf{$^{226}$Ra} & \textbf{$^{232}$Th} &\textbf{$^{228}$Th} & \textbf{$^{235}$U}& \textbf{$^{40}$K}  & \textbf{$^{60}$Co}& \textbf{$^{137}$Cs}\\
\hline
\endhead
&(Follows at next page)\\
\endfoot
\endlastfoot

1 & OFE Cu, Luvata  &  GDMS &mBq/kg& $<$0.012&& $<$0.0041&&& 0.061&&\\
2 & Kapton-Cu, LabCircuits  &  Ge & $\mu$Bq/cm$^{2}$ & $<$160 & $<$14 & $<$12 & $<$8  & $<$2 & $<$40 & $<$2 & $<$2  \\
3 & Epoxy Hysol, Henkel  &  Ge & mBq/kg & $<$273 & $<$16 &$<$20& $<$16 & & $<$83& $<$4.2& $<$4.5 \\
4 & SM5D resistor, Finechem  &  Ge  & mBq/pc& 0.4$\pm$0.2 & 0.022$\pm$0.007 & $<$0.023 & $<$0.016  & 0.012$\pm$0.005 & 0.17$\pm$0.07& $<$0.005 & $<$0.005 \\
5 & Connectors, Fujipoly  &  Ge & mBq/pc & $<$25 & 4.45$\pm$0.65 & 1.15$\pm$0.35 & 0.80$\pm$0.19 & & 7.3$\pm$2.6 & $<$0.1 & $<$0.4\\
6 & Flat cable, Somacis  &  Ge & mBq/pc & $<$14 & 0.44$\pm$0.12 & $<$0.33 & $<$0.19 & $<$0.19 & 1.8$\pm$0.7 & $<$0.09 & $<$0.10 \\
7 & Teflon cable, Druflon  &  Ge & mBq/kg & $<$104 & $<$2.2 & $<$3.7 & $<$ 1.7 & $<$1.4 & 21.6$\pm$7.4 & $<$0.7 & $<$0.8 \\
8 & Coaxial cable, Axon &  Ge & mBq/kg & $<$650 & $<$24 & $<$15 & $<$9.9 & $<$7.9 & 163$\pm$55 & $<$4.3 & $<$5.1 \\
9 & Classical Micromegas, CAST &  Ge& $\mu$Bq/cm$^{2}$ & $<$40 &&4.6$\pm$1.6&& $<$6.2& $<$46 & $<$3.1& \\
10 & Microbulk Micromegas, CAST &  Ge & $\mu$Bq/cm$^{2}$ & 26$\pm$14&& $<$9.3&& $<$14 &57$\pm$25& $<$3.1& \\
11 & Kapton-Cu foil, CERN  &  Ge& $\mu$Bq/cm$^{2}$ & $<$11&& $<$4.6 &&$<$3.1& $<$7.7 &$<$1.6&\\
12 & Cu-kapton-Cu foil, CERN  &  Ge& $\mu$Bq/cm$^{2}$ & $<$11 && $<$4.6&& $<$3.1& $<$7.7& $<$1.6& \\
13 & Vacrel foil, Saclay  &  Ge& $\mu$Bq/cm$^{2}$& $<$19& $<$0.61&$<$0.63& $<$0.72& $<$0.19& 4.6$\pm$1.9& $<$0.10 &$<$0.14 \\
14 & Kapton-diamond foil, CERN &  Ge & $\mu$Bq/cm$^{2}$ & $<$350 & $<$13 & $<$14 & $<$11 & $<$6 & $<$51 & $<$2 & $<$3 \\
15 & Ceramic, Saclay  &  Ge& Bq/kg & 6.0$\pm$3.2 & 0.52$\pm$0.07 & 0.18$\pm$0.05 & 0.15$\pm$0.02& 0.38$\pm$0.09 & 2.0$\pm$0.6 & $<$0.03 & $<$0.05 \\ \hline \hline
\textbf{\#} & \textbf{Material,Supplier} &   \textbf{Method} &\textbf{Unit} & \textbf{$^{214}$Bi} &  & \textbf{$^{208}$Tl} & & &   & & \\ \hline
16 & Microbulk Micromegas, CAST/CERN &  BiPo-3  & $\mu$Bq/cm$^{2}$ & \multicolumn{2}{c}{$<$ 0.134 } & $<$ 0.035 &&  && & \\
17 & Cu-kapton-Cu foil, CERN  &  BiPo-3 & $\mu$Bq/cm$^{2}$ & \multicolumn{2}{c}{$<$ 0.141}  & $<$0.012&& & & & \\
18 & Kapton-epoxy foil, CERN  &  BiPo-3 & $\mu$Bq/cm$^{2}$ & \multicolumn{2}{c}{$<$ 0.033} & $<$0.008 &&&  &&\\
19 & Vacrel foil, Saclay  &  BiPo-3 & $\mu$Bq/cm$^{2}$ & \multicolumn{2}{c}{$<$ 0.032} & $<$0.013 &&&  &&\\
20 & Kapton-diamond foil, CERN &  BiPo-3 & $\mu$Bq/cm$^{2}$ & \multicolumn{2}{c}{$<$ 0.055} & $<$0.016 &&&  &&\\

\hline \caption{Activities measured for the most relevant samples analyzed in the radiopurity assessment program carried out for T-REX. Most results were obtained by germanium spectrometry, except for those at \#1, derived from GDMS, and \#16-20, obtained with the BiPo-3 detector (see text for details). Values reported for $^{238}$U and $^{232}$Th correspond to the upper part of the chains and those of $^{226}$Ra and $^{228}$Th give activities of the lower parts. Reported errors correspond to $1\sigma$ uncertainties and upper limits are evaluated at 95\% CL. For the samples \#16-20, values for the $^{214}$Bi and $^{208}$Tl isotopes are given and limits correspond to 90 \% C.L.}
\label{tab:radiopurity}
\end{longtable}
\end{landscape}
\normalsize


\section{Operation of Micromegas in Xenon and TMA mixtures}
\label{sec:operation}


It is well known that charge amplification in pure noble gases is problematic due to the rapid photon-driven expansion of the avalanche, which makes the detector quickly depart from the proportional amplification regime into the ``sparking'' Geiger regime. This is the main reason of the use of quenchers in conventional gas TPCs, gas additives that are able to effectively reduce or \textit{quench} the photon production in the avalanche. For a DBD experiment, the use of the scintillating properties of the pure Xe are motivated, e.g., to obtain the time stamp of the event ($t_0$) and consequently accurate fiducialization in the drift direction $z$. It is remarkable that microbulk Micromegas can operate at high pressure pure Xenon, even with modest gains~\cite{Cebrian:2010nw,Balan:2010kx,tesisdiana}, something that is usually not possible with other MPGDs. Gains of around 500~(100) have been demonstrated in small HPXe setups at 1(10) bar, as well as energy resolution of around 12(30)\% FWHM for the 22 keV peak. This result is attributed to the fact that the confinement of the avalanche in microbulk readouts (inside the kapton cell formed below each micromesh hole) prevents the photons from expanding the avalanche far away, and acts as a sort of geometrical quencher.

This result seems to point that \textit{in principle} a Micromegas-TPC in pure Xe is feasible. However, as will be argued in the following, the addition of an adequate quencher to the Xe brings important advantages that largely compensate the absence of scintillation. Although keeping some scintillation in Xe with a quencher (with, maybe, CF$_4$~\cite{Broggini1994503}) is not excluded, we focus here on a pure charge-readout implementation (i.e. a $t_0$-less TPC -- see later on discussion in section~\ref{sec:discussion}) in Xenon with trimethylamine (TMA). This additive has been shown to be particularly beneficial for operation of Micromegas in Xenon. TMA forms a Penning mixture with Xe and this translates to higher gain at the same voltage, higher maximum gains, and better energy resolution. As discussed below, Xe+TMA turns out to enjoy extremely low electron diffusion, a very appealing feature for DBD searches.

\begin{figure}[t!]
\centering
\includegraphics[height=6cm]{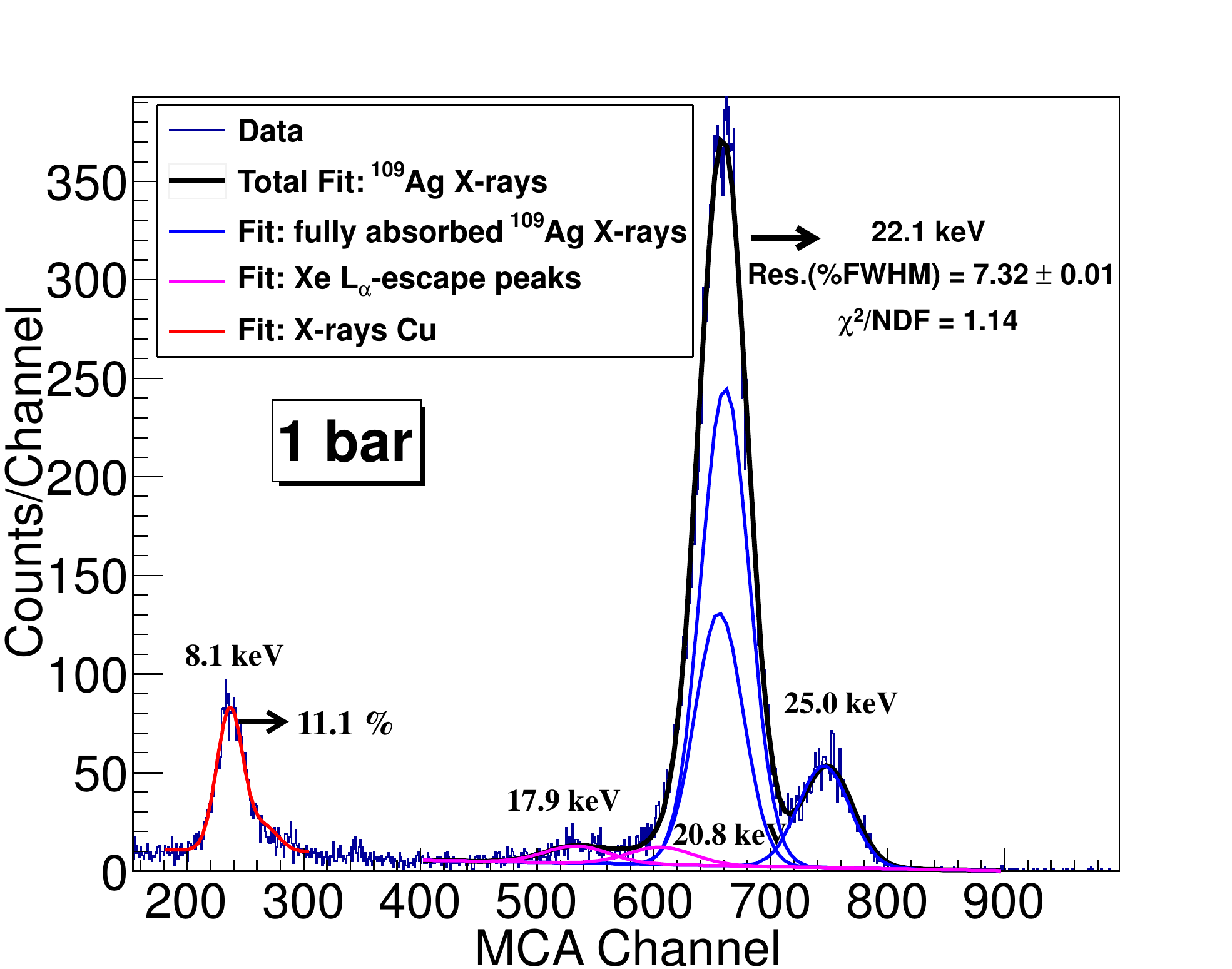}
\includegraphics[height=6cm]{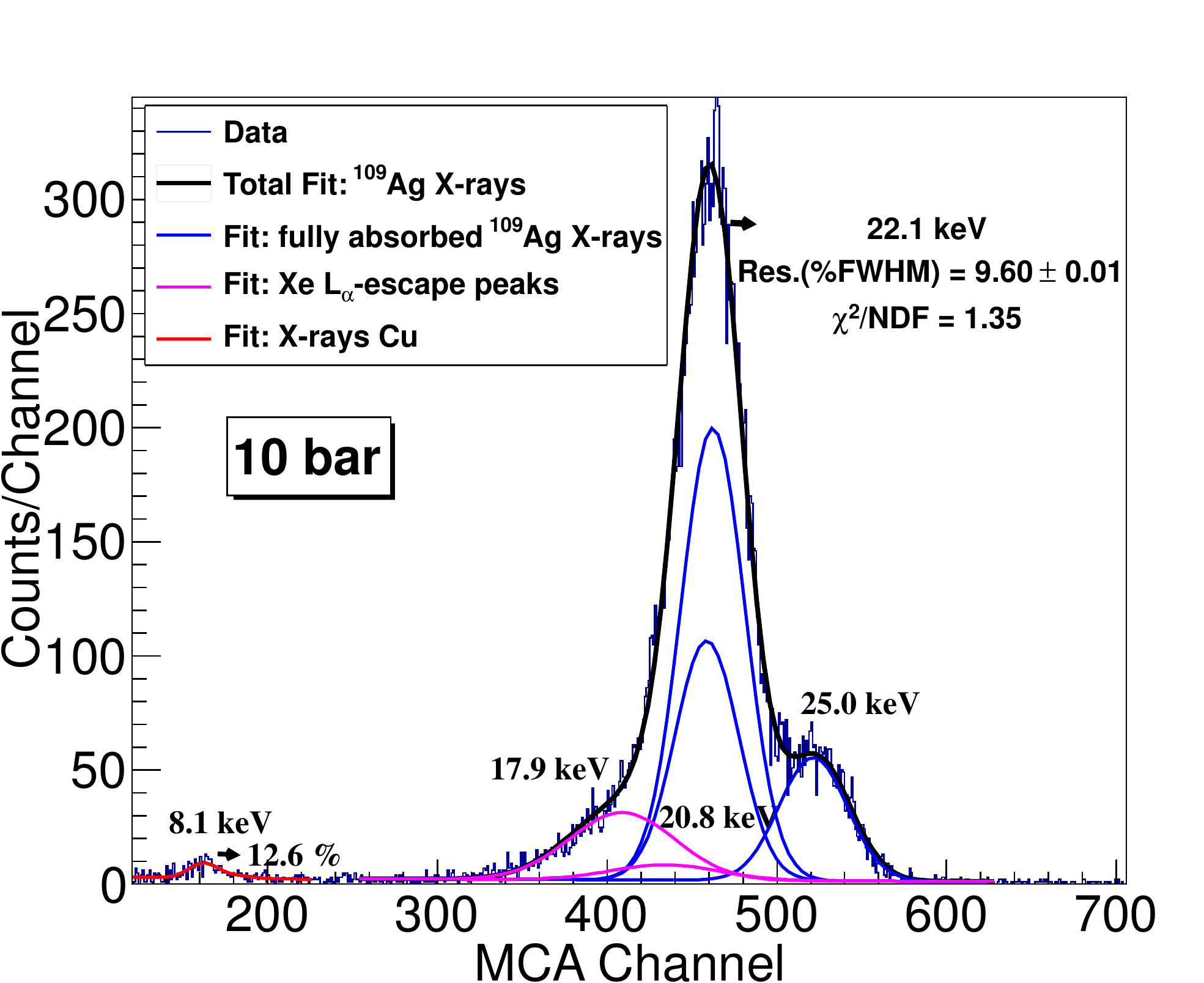}
\caption{Energy spectrum acquired with a $^{109}$Cd source in a 99.38\%-0.68\% Xe-TMA mixture at 1~bar (left) and at 10~bar (right).
The characteristic peaks at 22.1~keV and 25~keV are clearly seen, as well as the Xe escape peak of the former.
The overall fit is taking into account both of the Xe escape peaks. At the lower energy part,
the Cu K-fluorescence peak at 8.1~keV is also visible. The energy resolution measured for the 22.1 keV peak is 7.3\% FWHM and 9.6~\% FWHM respectively. Figures taken from \cite{Cebrian:2012sp}.}
\label{fig:spectrum}
\end{figure}

\begin{figure}[htb!]
\centering
\includegraphics[width=0.48\textwidth]{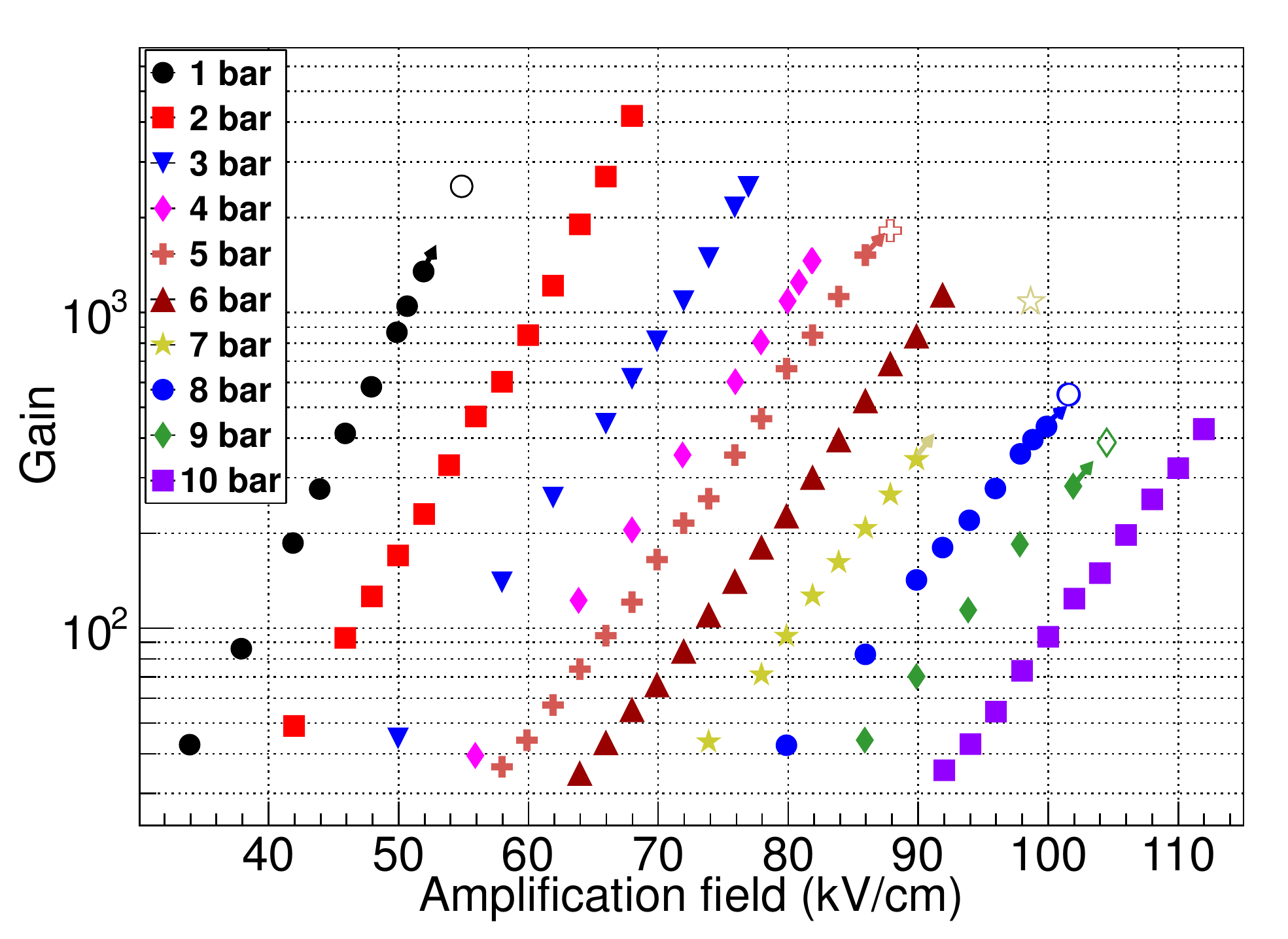}
\includegraphics[width=0.48\textwidth]{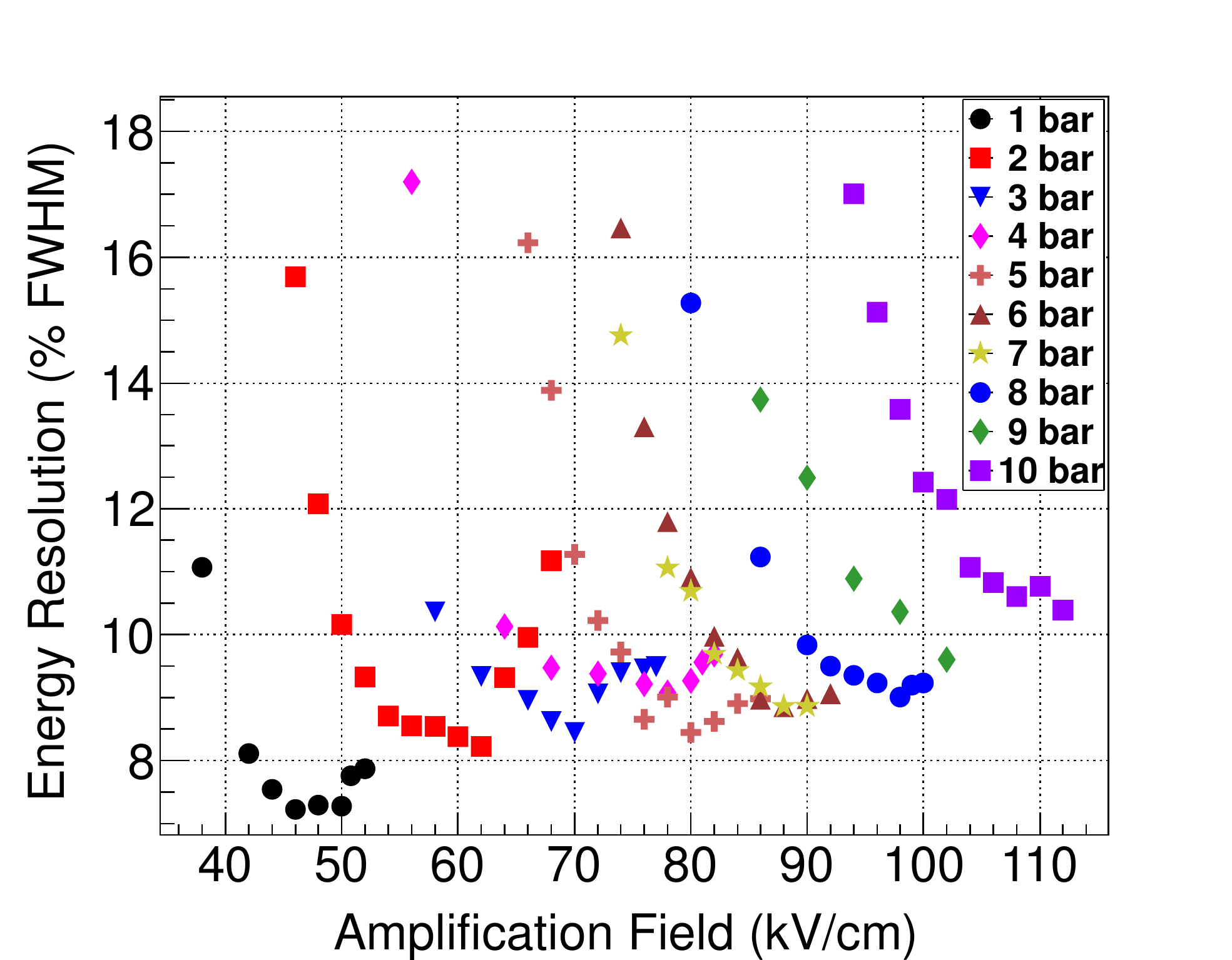}
\caption{Gain (left) and energy resolution (right) at the 22.1 keV peak versus on the amplification field
for operating pressures between 1~bar and 10~bar, in optimal Xe-(0.85\%--1.6\%)TMA mixtures.
See \cite{Cebrian:2012sp} for more details.}
\label{fig:gaineres}
\end{figure}

As part of the T-REX project an exhaustive program of characterization of this mixture has been carried out, with varying concentrations of TMA and at pressures from 1~bar to 10~bar \cite{Cebrian:2012sp}. Data were taken in a small 2.4 l high-pressure high-purity chamber equipped with a single-anode 3-cm diameter microbulk, using low energy gammas from $^{241}$Am and $^{109}$Cd calibration sources. Figure \ref{fig:spectrum} shows a typical $^{109}$Cd energy spectrum at 10 bar, where a fit to the two prominent peaks (at 22.1~keV and at 25~keV) of the source and the corresponding escape peaks from Xe is included.

Maximum gains are achieved, generally, for TMA concentrations of the order of $\sim 1\%$, with maximum
values better than 2000~(400) for 1~(10)~bar. Apart from the higher gain with respect to operation in pure Xe, the stability in the operation of the Micromegas (e.g. reduced number of discharges) is notable.
The measurements at 1~bar have yielded an energy resolution of 7.3\%~(FWHM) at 22.1~keV for TMA concentrations above 0.9\%. At higher pressures, the best energy resolution obtained for the same peak is 8.3\%, 9.0\% and 9.6\% (FWHM) at 5, 8 and 10~bar respectively, with TMA concentrations within the range of  0.7\%--1.6\% (see Figure~\ref{fig:spectrum}). These values are a factor of $\sim$3 better than the ones obtained in pure Xe. Figure~\ref{fig:gaineres} gathers a large number of measurements of gain and energy resolution for different amplification voltages and gas pressure.


Summarizing, the operation of Micromegas in Xe-TMA, with TMA concentrations within the range of  0.7\%--1.6\%, enjoys higher gain for the same amplification field, better stability and energy resolution, with respect to that of pure Xe. In general, the performance of the Micromegas is comparable to optimum values in benchmark Micromegas mixtures (e.g. optimum Ar+Isobutane mixtures). The values of energy resolution presented above, if extrapolated to the $^{136}$Xe \Qbb\ energy following a $1/\sqrt{E}$ law, correspond to 0.7\% and 0.9\% FWHM, for 1 and 10 bar respectively. These values should be achievable in ideal conditions, but they must be realistically measured with long electron high-energy tracks of comparable energy as the DBD events (see section~\ref{sec:etracks}). However, they prove  that sub-\% resolution at \Qbb\ could \textit{in principle} be achieved with Micromegas readouts and high pressure Xe-TMA gas.

\begin{figure}[t!]
\centering
\includegraphics[width=0.7\textwidth]{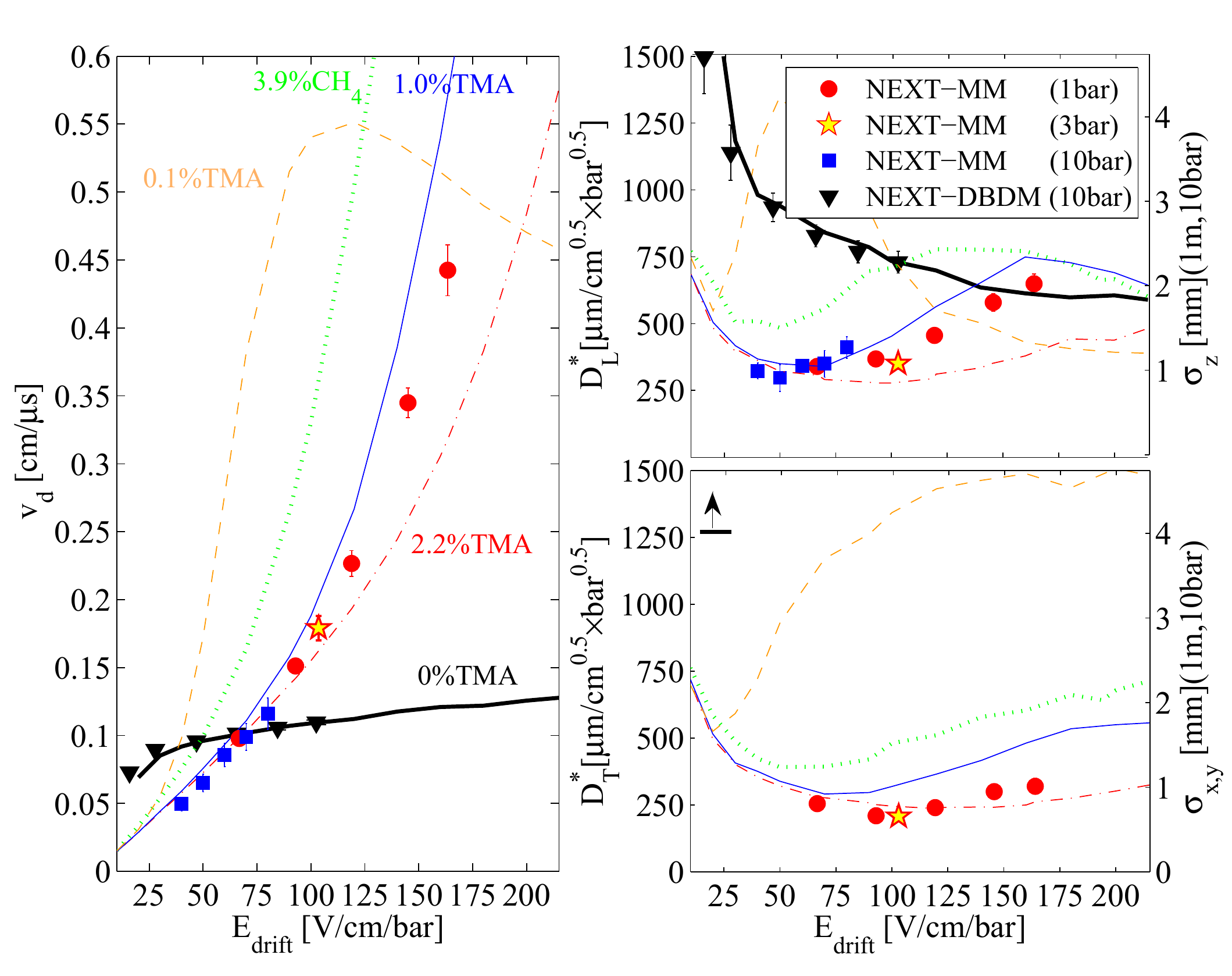}
\caption{Drift velocity (left) and longitudinal and transversal diffusion (right)
in Xe-TMA mixtures compared with other gas mixtures.
The dots correspond to experimental data and the lines
to microscopic simulations done with Magboltz. We refer to \cite{GonzalezDiaz20158} for further details on the measurements.}
\label{fig:diff}
\end{figure}

The addition of TMA to the Xe improves operation in another very important aspect. As part of T-REX, electron diffusion and drift properties in Xe+TMA mixtures have been measured for the first time, for a number of gas parameters (pressure and TMA concentration)~\cite{GonzalezDiaz20158}. They turn out to be compatible with the values anticipated by microphysics simulations with Magboltz~\cite{Biagi:1999nwa}, as shown in Figure~\ref{fig:diff}. The drift velocity measured is similar to the one of pure Xe, of the order of 0.1~cm/$\mu$s, for the field values of interest (around 75~V/cm/bar). However, the electron diffusion parameters are substantially improved: values of 300~$\mu$m~cm$^{-1/2}$~bar$^{1/2}$ for the longitudinal diffusion and 250~$\mu$m~cm$^{-1/2}$~bar$^{1/2}$ for the transversal are measured for a drift field of 750~V/cm and 1\% of TMA at 10 bar. These values are a factor $\sim$20 and $\sim$3 better (respectively for the transversal and longitudinal diffusion coefficients) than the ones in pure Xenon, and indeed make this mixture one with the lowest electron diffusion known. They correspond to a \emph{blurring} at the scale of $\sim$1 mm, for events drifting 1~m. This feature of Xe+TMA mixtures is an important asset that will impact topological information.

%
%
%


\section{Discrimination by topological pattern recognition of DBD events}
\label{sec:topology}

\begin{figure}[b!]
        \centering
                \includegraphics[height=4.4cm]{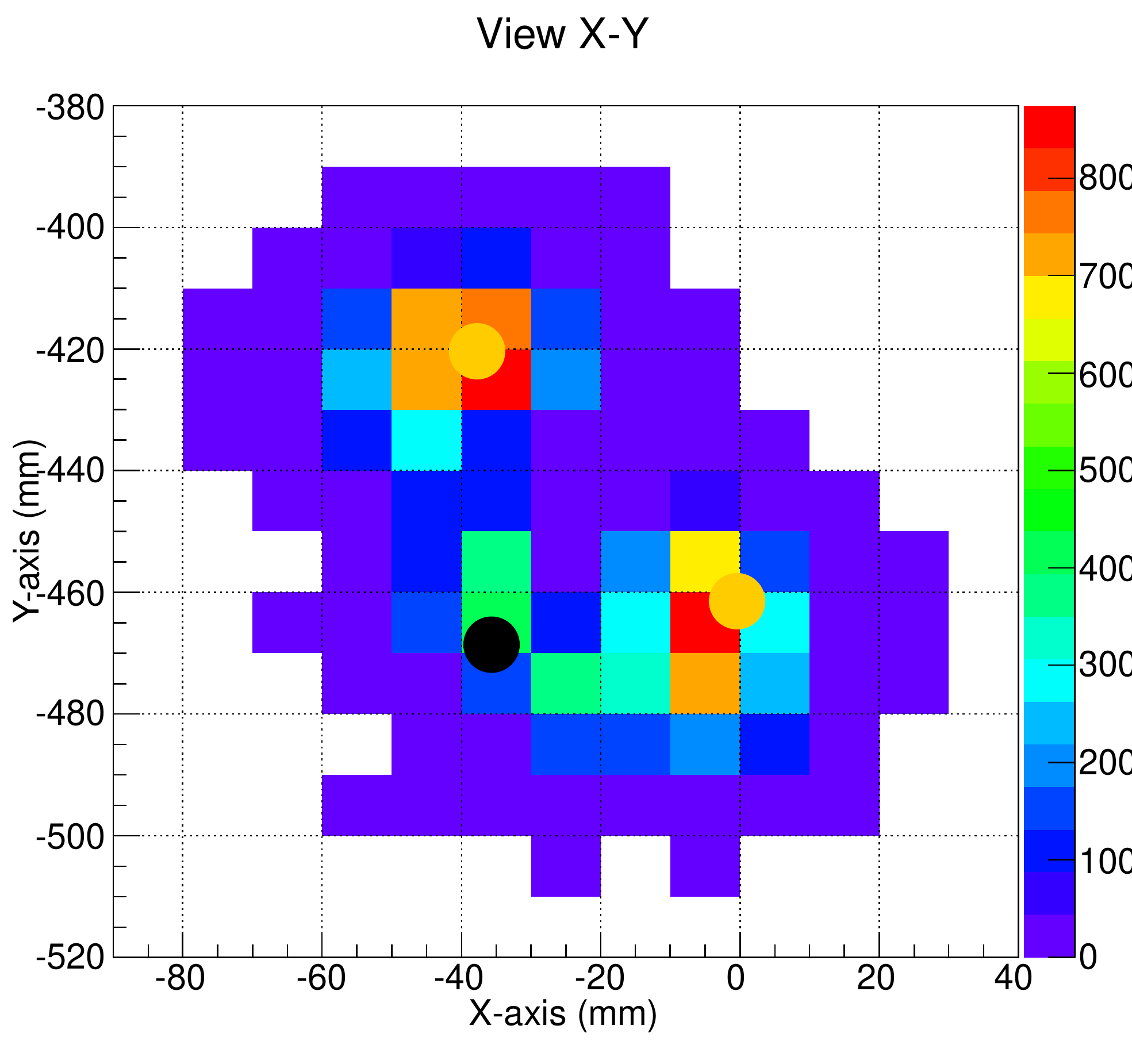}
                \includegraphics[height=4.4cm]{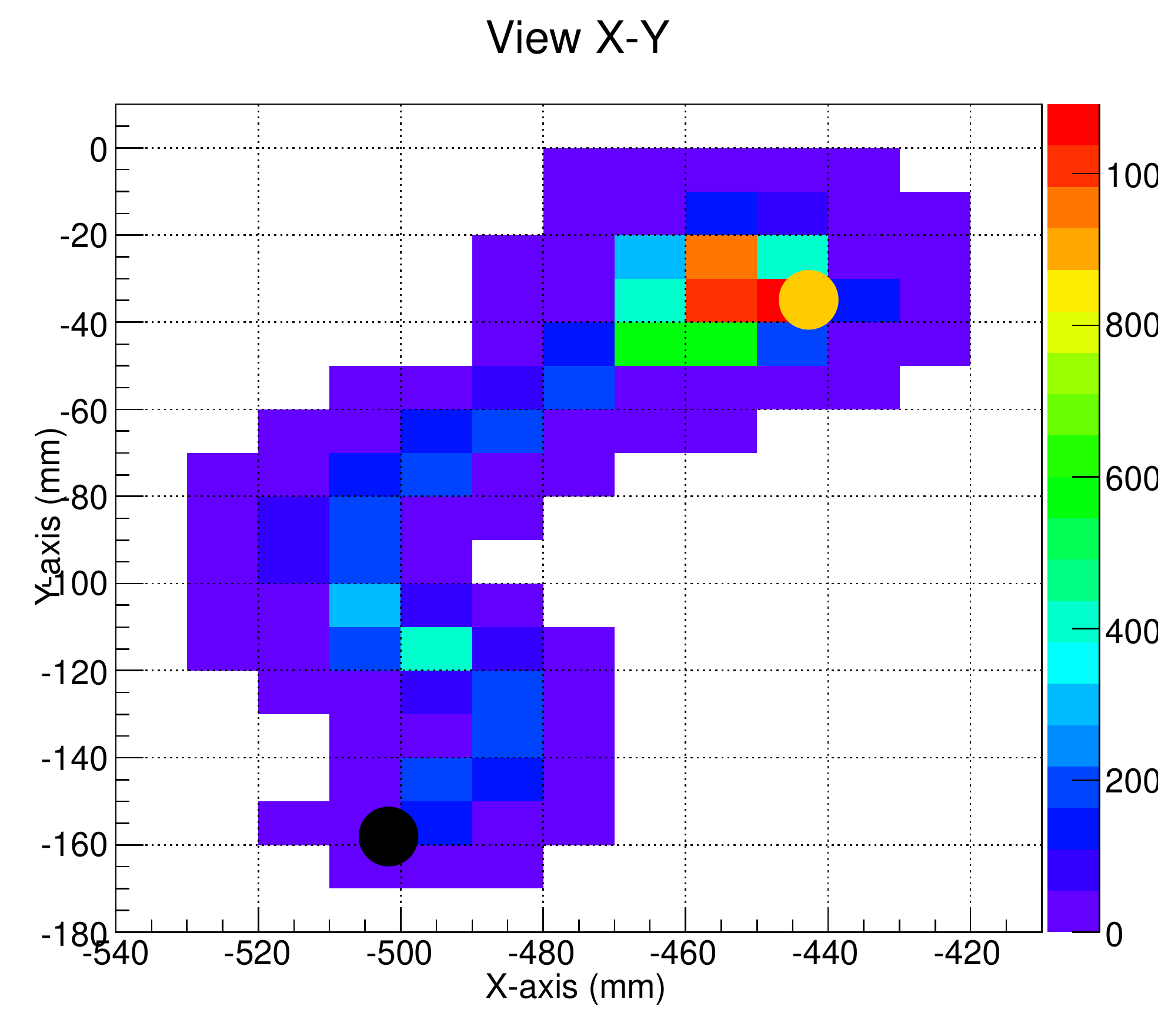}
                \includegraphics[height=4.4cm]{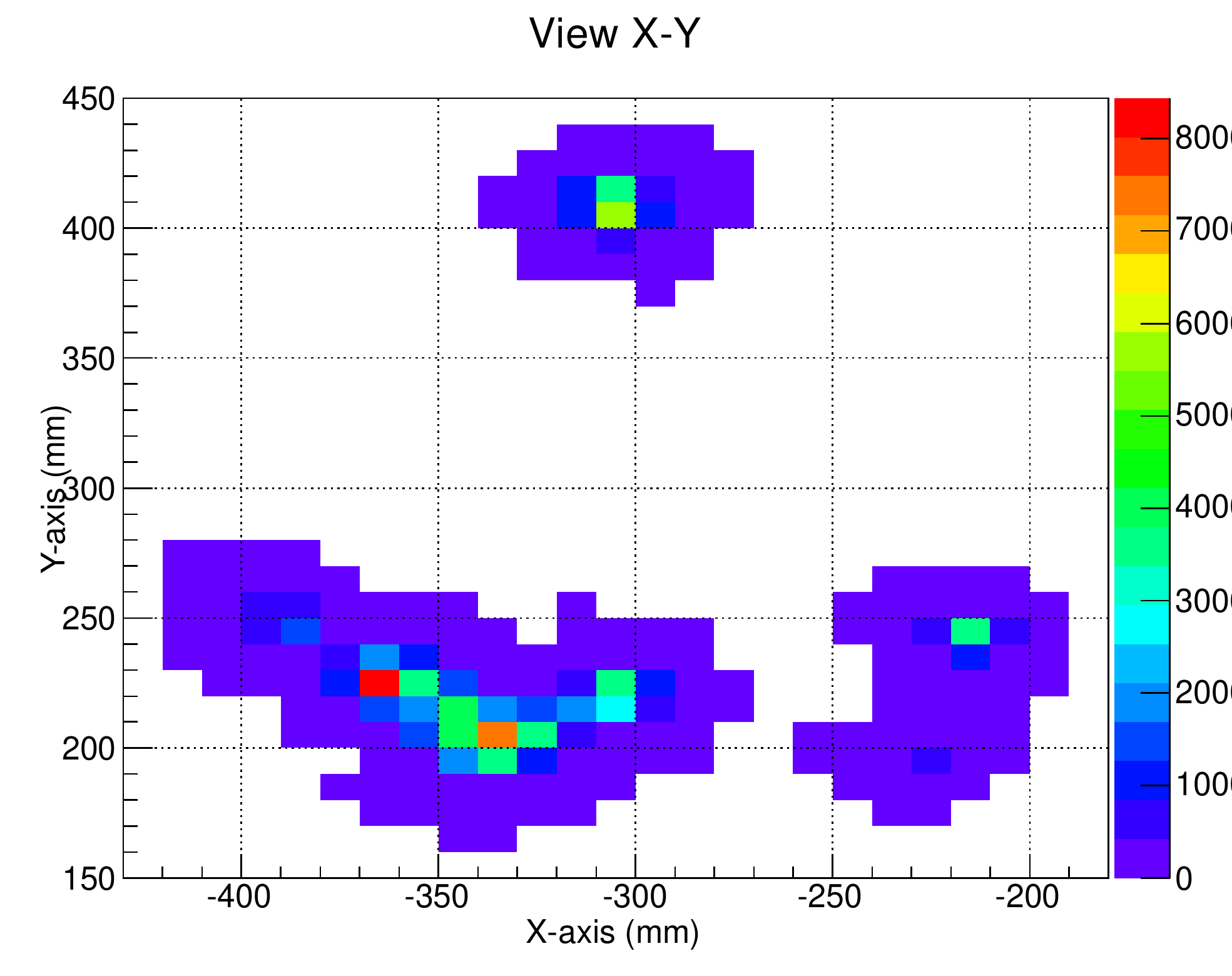}
        \caption{On the left, a 2D projection of a $^{136}$Xe $\beta\beta0\nu$ event simulated in 10\,bar Xenon with a relatively coarse pixelization of 10$\times$10 mm. In the middle, a $^{214}$Bi electron coming out from the lateral wall, and on the right, an multi-track event caused by a $^{208}$Tl 2614.5\,keV photon}\label{fig:events}
\end{figure}

\begin{figure}[t!]
\centering
\includegraphics[width=0.45\textwidth]{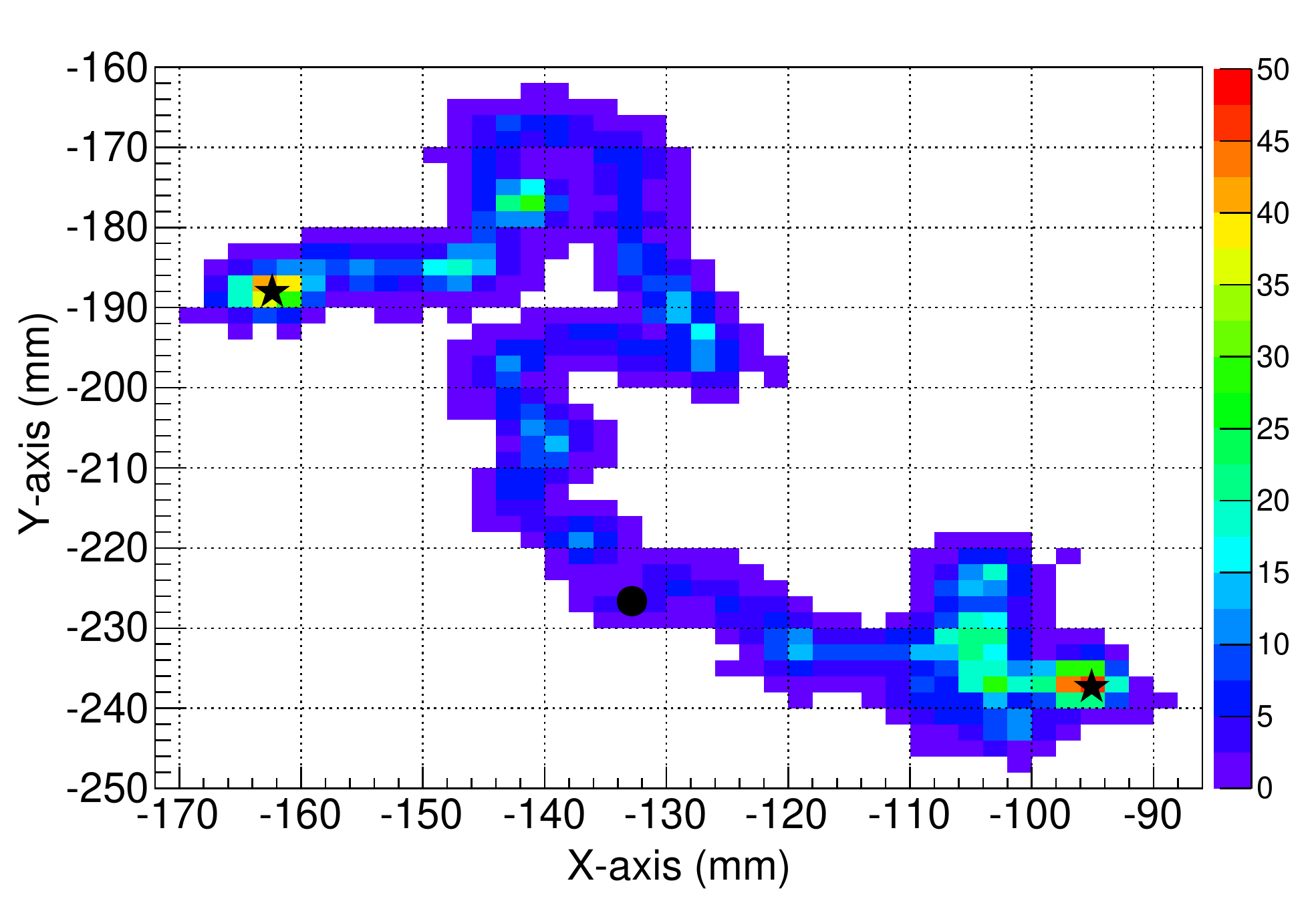}
\includegraphics[width=0.45\textwidth]{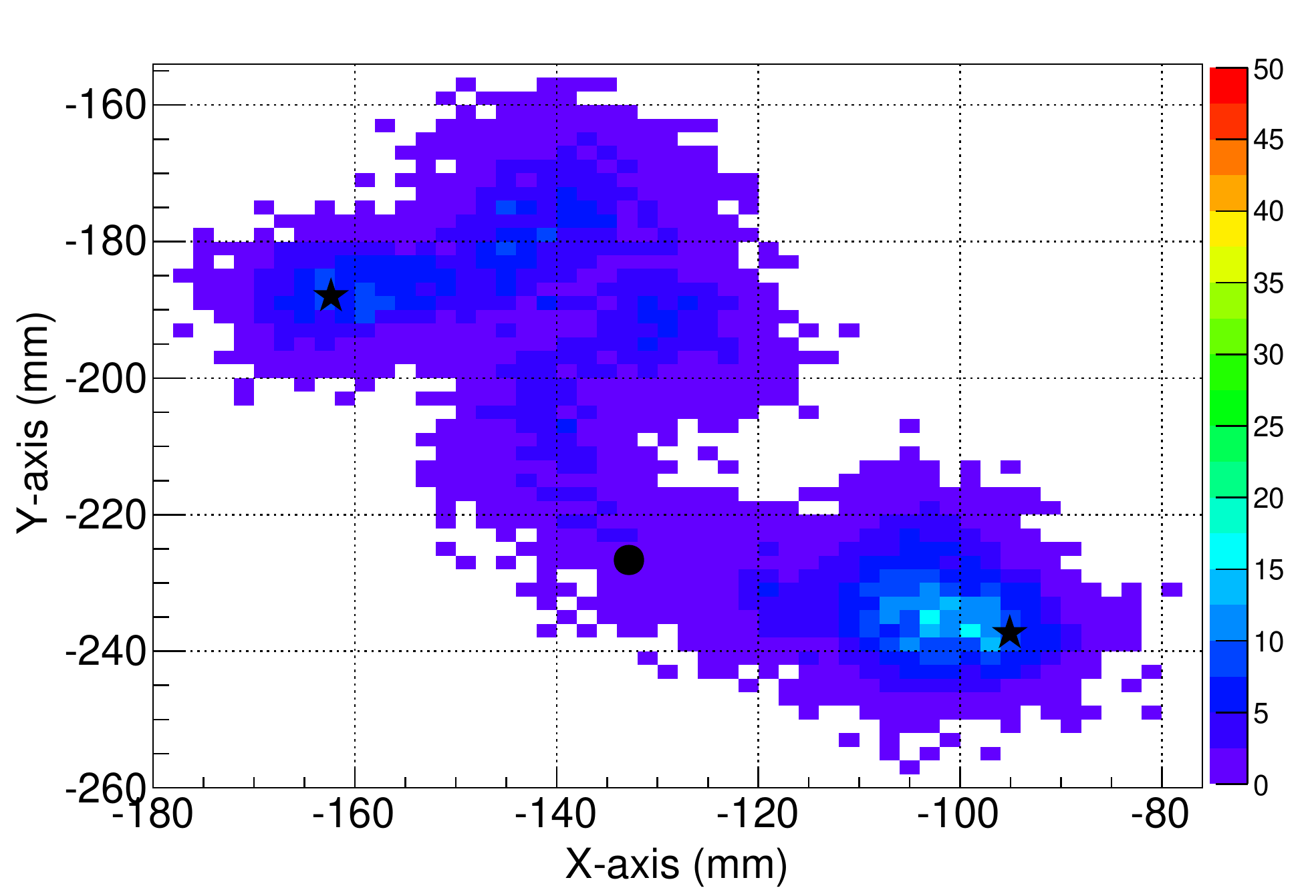}
\caption{A simulated $^{136}$Xe $\beta\beta0\nu$ event HPXe at 10 bar with a relatively fine readout granularity of 2$\times$2 mm pixels, after having drifted $\sim$50~cm in a low diffusion mixture like Xe+2\%TMA (left) and in a pure Xenon (right), supposing a drift field of 300~V/cm/bar and using the gas parameters of Magboltz \cite{Biagi:1999nwa}.
The color scale indicates the pixel's energy in keV. The event consists in two electrons emitted from the event vertex indicated with a black circle. The two electron blobs are marked by black stars.}
\label{fig:TopXe136Diff}
\end{figure}

The most powerful feature of gas TPCs for DBD searches is the possibility to image the DBD ionization topology in the gas, and use  it to distinguish it from background events. It translates into an extra background rejection factor that is out of reach of other detection techniques. It also represents a useful handle to study and identify possible background sources, and, very importantly, to  eventually confirm the DBD nature in the case of a putative positive signal. The typical topology of a $^{136}$Xe DBD event consists of a single straggling track produced by the two back-to-back electrons, ending in two larger energy depositions, or \emph{blobs}, at both ends of the track, corresponding to the Bragg peak of the electrons as they slow down to rest (see Figures~\ref{fig:events} and \ref{fig:TopXe136Diff}). The typical length of DBD events in 10 bar Xe is of 10--20 cm. The competing background topologies are those produced by high energy gammas (mainly 2614.5 keV photons from the decay of $^{208}$Tl, and 2447.8 keV photons from the $^{214}$Bi decay). They may present multi-track topologies (in case of one or more Compton interactions in the detector volume) or, alternatively, a single long straggling track ending in \emph{only one} blob.  A priori, the difference between these two topologies is accesible by a properly pixelized TPC, as demonstrated by the Gotthard group \cite{Wong:1993uq} long ago. Of course, these are \emph{prototype} topologies. In reality, a fraction of background events may mimic signal events and viceversa, e.g. DBD events may present multi-track topologies, due to bremsstrahlung, and background topologies may present extra (fake) blobs due to a particularly convoluted straggling of the electron track. A detailed study of the real discrimination capabilities under the light of a Micromegas-TPC implementation like the one here discussed has been one of the goals of the T-REX project.

First results in this direction~\cite{pacotesis,lauratesis,Cebrian:2013mza} have focused on the definition of an automated pattern recognition algorithm that could extract the needed distinguishing features out of the registered pixelised image of the events. The rationale of the algorithm, as presented in \cite{Cebrian:2013mza}, is based on: 1) identification of the number and charge of tracks and selection of those that are single tracks; 2) identification of the number and charge of the blobs in the main track of the event and select those with two blobs; and 3) fiducialization of the event so that surface or partially-contained events are rejected. To this basic skeleton, a number of tuning sub-rules are added. We refer to \cite{Cebrian:2013mza} for the specifics on the algorithm.

This algorithm has been tested on a representative sample of simulated DBD and background (both $^{208}$Tl and  $^{214}$Bi decays) events, in a generic 100-kg 10-bar Xe cylindrical volume. Its performance rejecting background events depends on the geometrical origin of the background contamination. Therefore, the study is performed with four populations of background events, representative of the most relevant geometrical regions. They are labelled as: ``vessel'', for events generated volumetrically in a thick copper-based vessel surrounding the active volume; ``field-cage'', for events generated on a surface directly bounding laterally the active volume; ``readout'', for events generated on a surface directly touching the active volume at the anode of the TPC (e.g. on the Micromegas readouts); and ``cathode'', for events generated on a surface at the cathode of the TPC.

The reduction factor obtained in \cite{Cebrian:2013mza} when applying the algorithm to background events ranges from (0.2--7) $\times 10^{-6}$ depending on the place and type of the contamination. Table \ref{tab:AccfacTotXe} shows the factors obtained for each of the sample populations independently. This factor is expressed with respect to the total number of decays simulated, and therefore includes a $\sim10^{-3}$ factor (see third column of table \ref{tab:AccfacTotXe}) resulting just from the probability that a given decay produces an energy deposit in the energy region of interest (RoI) of 3\% around $Q_{\beta\beta}$. Therefore, topological cuts alone can reduce background in the RoI by an additional factor 10$^{-2}$-10$^{-3}$ depending on the type of background. Compared with a simple single-site/multi-site rejection capability (approximately representative, e.g. of what liquid Xenon experiments can provide), the full topological recognition in gas (i.e. adding blob identification) gives an extra rejection factor of $\sim$10. When applied to signal events (see the last row of Table~\ref{tab:AccfacTotXe}), the algorithm shows a signal efficiency of 40\% (part of the efficiency loss --about 70\%-- comes from fiducialization, and will be lower the larger the detector size). These numbers are considered conservative, as the study in~\cite{Cebrian:2013mza} was preliminary and had a number of limitations. We comment on this later on and present some additional results beyond these numbers. In any case, a number of important learnings have been extracted from this study:

\begin{table}[b!]
\centering
\begin{tabular}{ccrrr} \hline
	Origin & Isotope &   Events in RoI & Final events (HD) & Final events (LD)\\

   & & ($\times 10^{-3}$) & ($\times 10^{-6}$)  & ($\times 10^{-6}$) \\\hline

\multirow{2}{*}{vessel}& $^{208}$Tl              &0.4  							&7.0$\pm$0.6              &2.5$\pm$0.4 \\
 & $^{214}$Bi			 &0.03            							  &1.5$\pm$0.1               &0.5$\pm$0.1\\[1mm]	
 \multirow{2}{*}{field-cage}& $^{208}$Tl    &2.0    							  &13.4$\pm$0.1               &7.2$\pm$0.9 \\
  		 & $^{214}$Bi			 &0.2   		  &4.1$\pm$0.3               &1.7$\pm$0.1\\[1mm]
  									
 \multirow{2}{*}{readout}& $^{208}$Tl    &2.7    								&5.4$\pm$0.7               &2.1$\pm$0.5\\
  		& $^{214}$Bi	 &1.4    &0.9$\pm$0.2               &0.2$\pm$0.1\\[1mm]		

 \multirow{2}{*}{cathode} & $^{208}$Tl      &2.2   							  &7.4$\pm$0.8               &4.4$\pm$0.6 \\
  	& $^{214}$Bi							   &1.0   								 &1.1$\pm$0.2               &0.4$\pm$0.1\\ \hline \hline
     & & ($\%$) & ($\%$)  & ($\%$) \\ \hline
      All Xe volume & $^{136}$Xe  $0\nu\beta\beta$      &71.1 $\pm$ 0.2     &40.7 $\pm$ 0.2 				 &40.0 $\pm$ 0.2\\\hline
\end{tabular}
\caption{\label{tab:AccfacTotXe} Background and signal (last row) reduction factors for the different event populations studied in a RoI around the $Q_{\beta\beta}$ (first column), and after the application of the selection criteria as defined in \cite{Cebrian:2013mza} in the case of pure xenon (second column) and in the case of a low diffusion xenon mixture (last column). The indicated errors are from the statistics of simulation.}
\end{table}

\begin{itemize}
  \item Low electron diffusion in the gas is an important asset and has an impact on the quality of the topological pattern recognition. The study in~\cite{Cebrian:2013mza} was limited by a relatively coarse pixel size of 10$\times$10 mm (the reason being that it was originally aimed at topological studies in pure Xe). Even with this limitation, the algorithms still presented a factor $\sim$3 better rejection factor for low diffussion (like in Xe+TMA) than for high diffusion (like in pure Xe) events, for all types of background. The impact of low diffusion concentrated on the ability to identify the two-blob topology.
  \item Fiducial cuts have little effect on background rejection, with the only exception of surface $^{214}$Bi events. The reason is that this isotope is $\beta$-emitter, and fiducialization helps tagging the remaining events that might have survived other cuts. This observation is important in order to assess the benefit of having $t_0$ information. The price to pay for not having $t_0$ in terms of background rejection turns out to be very small, especially if $^{214}$Bi contamination in the cathode and anode are under control. A symmetric TPC configuration with shared cathode additionally mitigates this drawback (see section~\ref{sec:discussion} for a discussion on the subject of the benefit of $t_0$).
  \item A number of observables have been qualitatively identified, but not included in the study, that could provide additional discrimination capability (like track length, blob charge and $dE/dx$,...). This is particularly true in the case of a low diffusion gas like Xe+TMA.
\end{itemize}

\begin{figure}[t!]
\centering
\includegraphics[width=75mm]{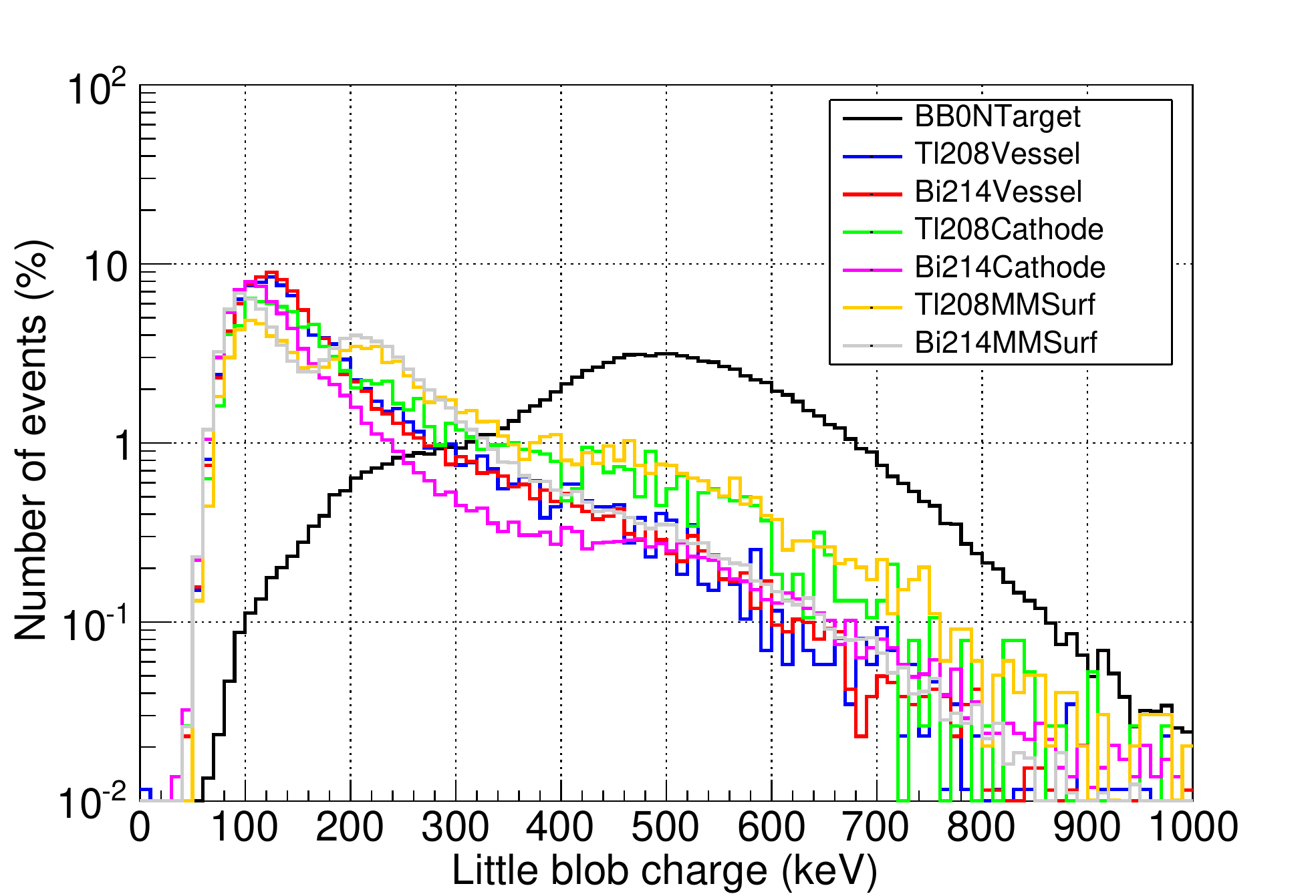}
\includegraphics[width=75mm]{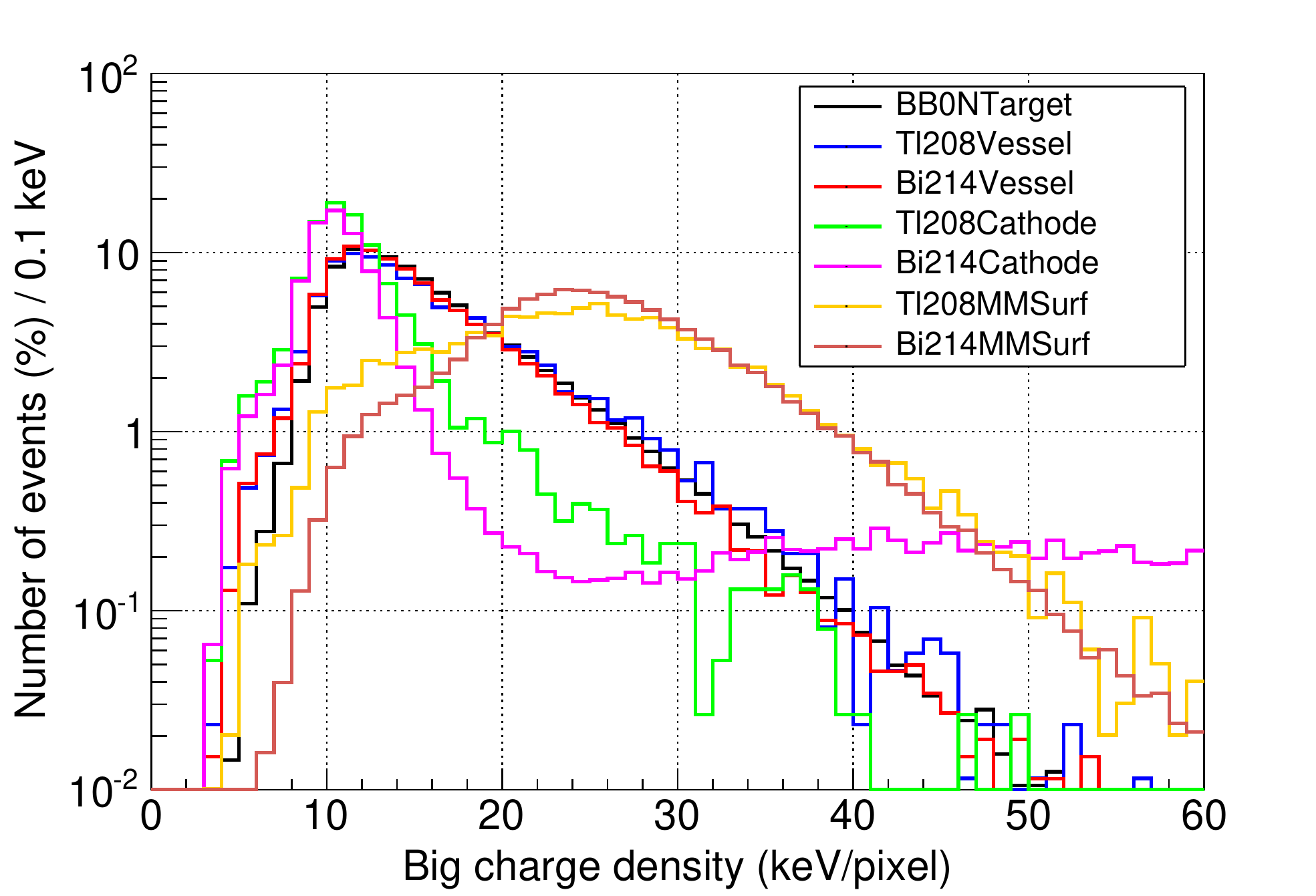}
\caption{ Distributions of the little blob charge (left) and the big blob charge density (right)
in a 200~kg high pressure TPC filled with Xe+1\%TMA at 10 bar for different populations:
$\beta\beta$0$\nu$ signal (black line),
$^{208}$Tl from the vessel (blue line), cathode (green line) and Micromegas detector (orange line);
and $^{214}$Bi from the vessel (red line), cathode (magenta line) and Micromegas detectors (grey line).
Events are in the RoI and have been previously selected by the track criterium.}
\label{fig:TopRejection}
\end{figure}

After the preliminary study in \cite{Cebrian:2013mza}, work has been focused in further exploring the topological capabilities of a low diffusion Xe+TMA gas with a more appropriate pixelization. Figure \ref{fig:TopXe136Diff} shows the same $^{136}$Xe $\beta\beta0\nu$ event with a simulated diffusion equivalent to 50~cm of drift in a low diffusion mixture like Xe+TMA, compared to pure Xe. The importance of low diffusion (and correspondingly good readout granularity) in the topological quality is evident to the eye. It allows more subtle topological features to emerge that could increase the discrimination capabilities of the algorithm. Some examples of potentially useful observables are the blob charge and density, the track length or the $dE/dx$ along the track. A detailed study is ongoing that supersedes that of~\cite{Cebrian:2013mza}, but it is out of the scope of the present paper. Just for the sake of illustration, Figure~\ref{fig:TopRejection} shows how two such observables are distributed for different event populations. Signal and background event populations present substantially different distributions.  The blob charge density shown on the right of Fig.~\ref{fig:TopRejection} shows a correlation with the event $z$-coordinate (events from the ``cathode'' population present lower blob charge density that those of the ``readout''). This is expected due to electron diffusion, and it illustrates the possibility of extracting diffusion information (and therefore $z$-coordinate information) from the event topology. This is further discussed in section~\ref{sec:discussion}. The impact of these refined topological criteria on the overall background rejection capability is still to be quantified with precision. First preliminary results suggest, however, that additional factors beyond the study presented in~\cite{Cebrian:2013mza} and Table~\ref{tab:AccfacTotXe} of up to $\sim$10 for some particular background populations may be achieved. Of course, at these levels of readout granularity and low diffusion the study must realistically incorporate signal readout effects like, e.g., pixel threshold effects, electronic noise, and temporal shaping, etc. to assure that the discrimination algorithm is inmune to those. In addition, this level of granularity corresponds to a very high number of readout channels if every pixel is to be read independently. Schemes like X-Y pixel multiplexation will allow for a large reduction of readout channels while keeping the advantage of high granularity.

\begin{figure}[t!]
\centering
\includegraphics[width=75mm]{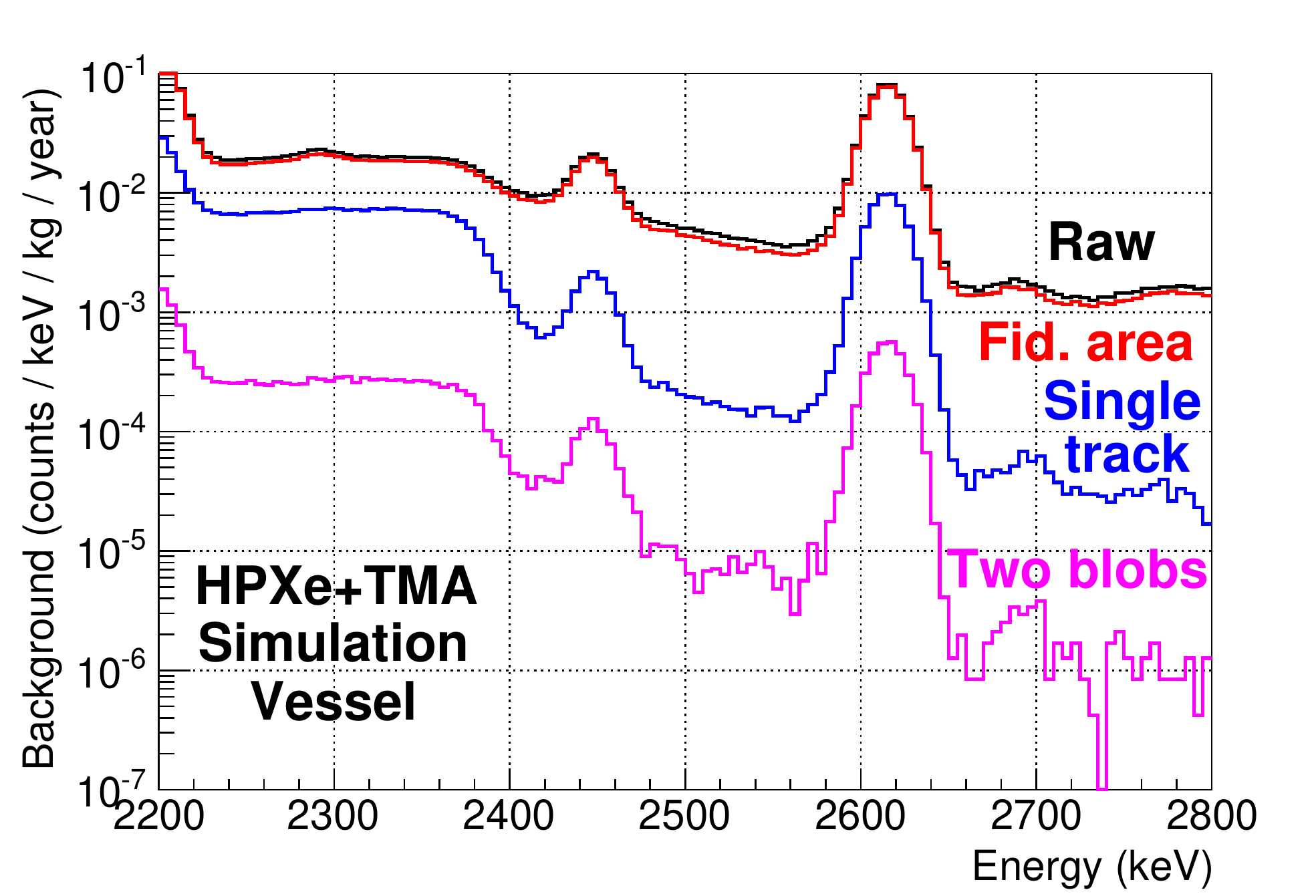}
\includegraphics[width=75mm]{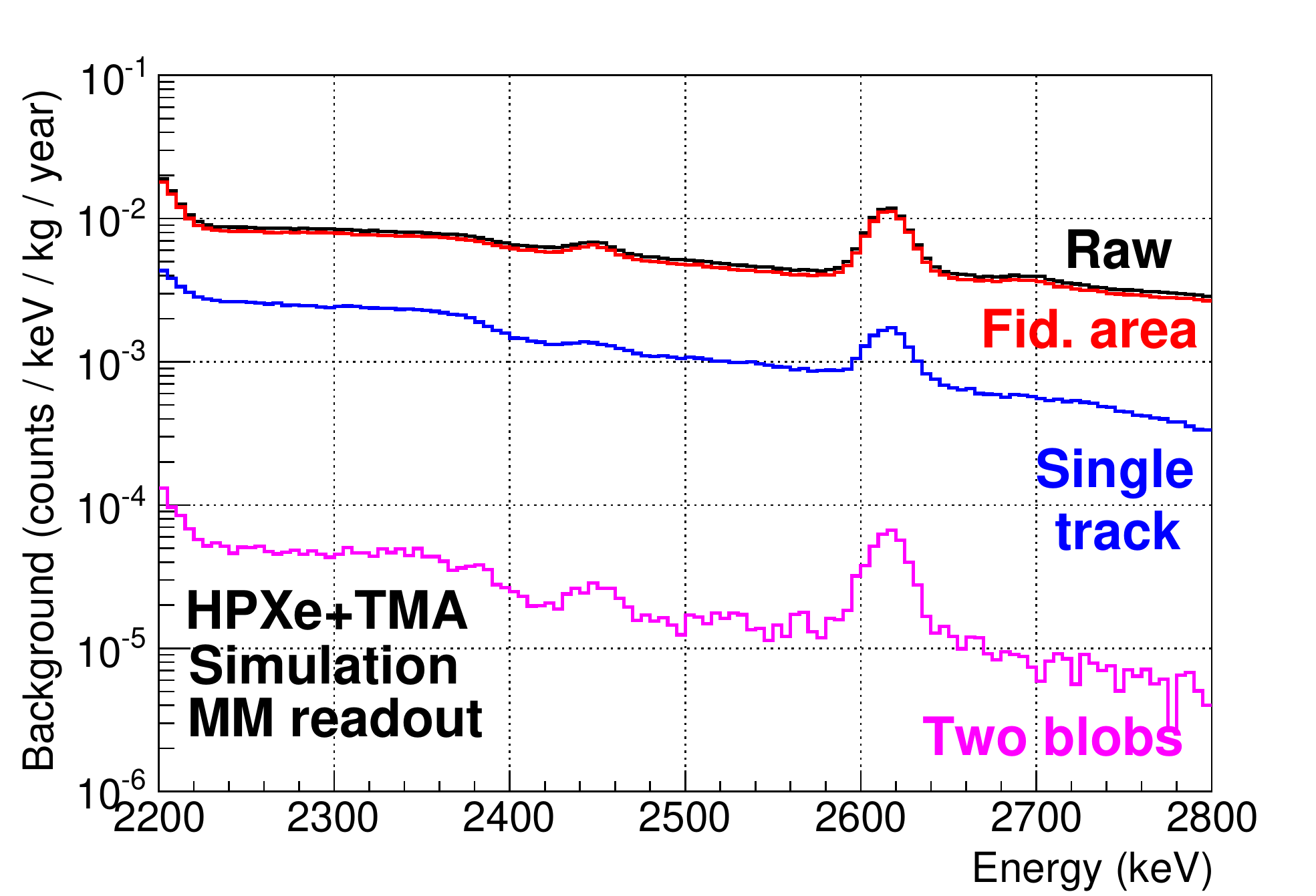}
\caption{Energy spectra from both $^{208}$Tl and $^{214}$Bi events simulated in
the ``vessel'' (left) and ``readout'' (right) geometries, in Xe+TMA and after the successive application of the selection criteria. Contrary to the results of Table~\protect\ref{tab:AccfacTotXe} and~\protect\cite{Cebrian:2013mza}, no fiducialization in $z$ is applied, while extra discrimination from the observables in Figure~\protect\ref{fig:TopRejection} is used. For both cases, the background events
are reduced to levels well below 10$^{-4}$ c~keV$^{-1}$~kg$^{-1}$~yr$^{-1}$.}
\label{fig:TopSpectra}
\end{figure}

Whether this discrimination capability results in competitive background levels also depends on the absolute radioactive levels present in the detector's environment. Therefore to make any precise statement on this question ones needs to build a complete background model including a detailed geometry of the final experiment and the contaminants associated with each of its materials and components. This task must be undertaken within the scope of a particular experiment, and is out of the scope of our study. However, it is possible to address the issue with some generality by applying the algorithm to representative radioactive populations associated with the detector configuration here advocated, under the light of the radiopurity results exposed in section~\ref{sec:micromegas}. We must note that it is a merit of this configuration that the materials and components that are close to the sensitive volume are reduced to a relatively simple set. Basically, the unavoidable levels are set by the microbulk foil at the anode, and the materials involved in the construction of the vessel and field cage, that can conceivably be reduced, e.g. to teflon and copper. No other aspect of the detector configuration is expected to set a radiopurity bottleneck for this approach (see section~\ref{sec:discussion} for additional discussion). Therefore, the illustrative populations of simulated events are normalized~\cite{Cebrian:2013mza} using generic volumetric and surface contamination levels (in reality, upper limits to those contaminations) associated to typical materials (e.g. copper and teflon for the vessel and field-cage, and the values obtained for the microbulk foils presented in section \ref{sec:micromegas} for the readout population). The background in the RoI thus obtained is at the level of, or below, 10$^{-4}$ c~keV$^{-1}$~kg$^{-1}$~yr$^{-1}$. This value corresponds to few counts of background in the RoI for typical exposures of a 100-kg scale experiment. Given that this value is fixed merely by upper limits to the radioactivity of the mentioned materials, and the still preliminary (and conservative) version of the discrimination algorithm in~\cite{Cebrian:2013mza}, even lower values might be reachable, maybe down to the requirements of the ton-scale experiments, as discussed later on.

\section{Imaging of long electron tracks with Micromegas}
\label{sec:etracks}

\begin{figure}[t!]
\centering
\includegraphics[height=75mm]{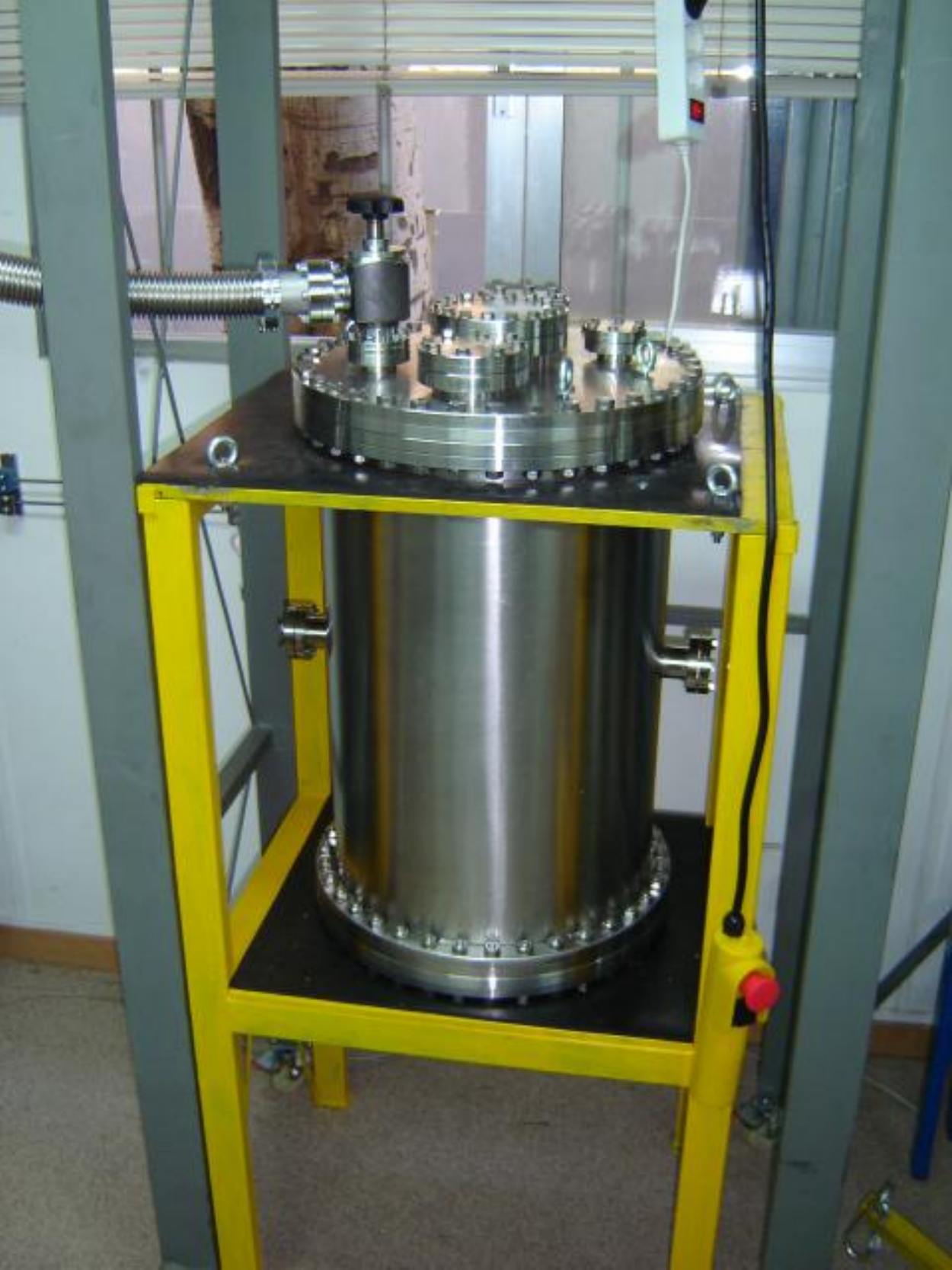}
\includegraphics[height=75mm]{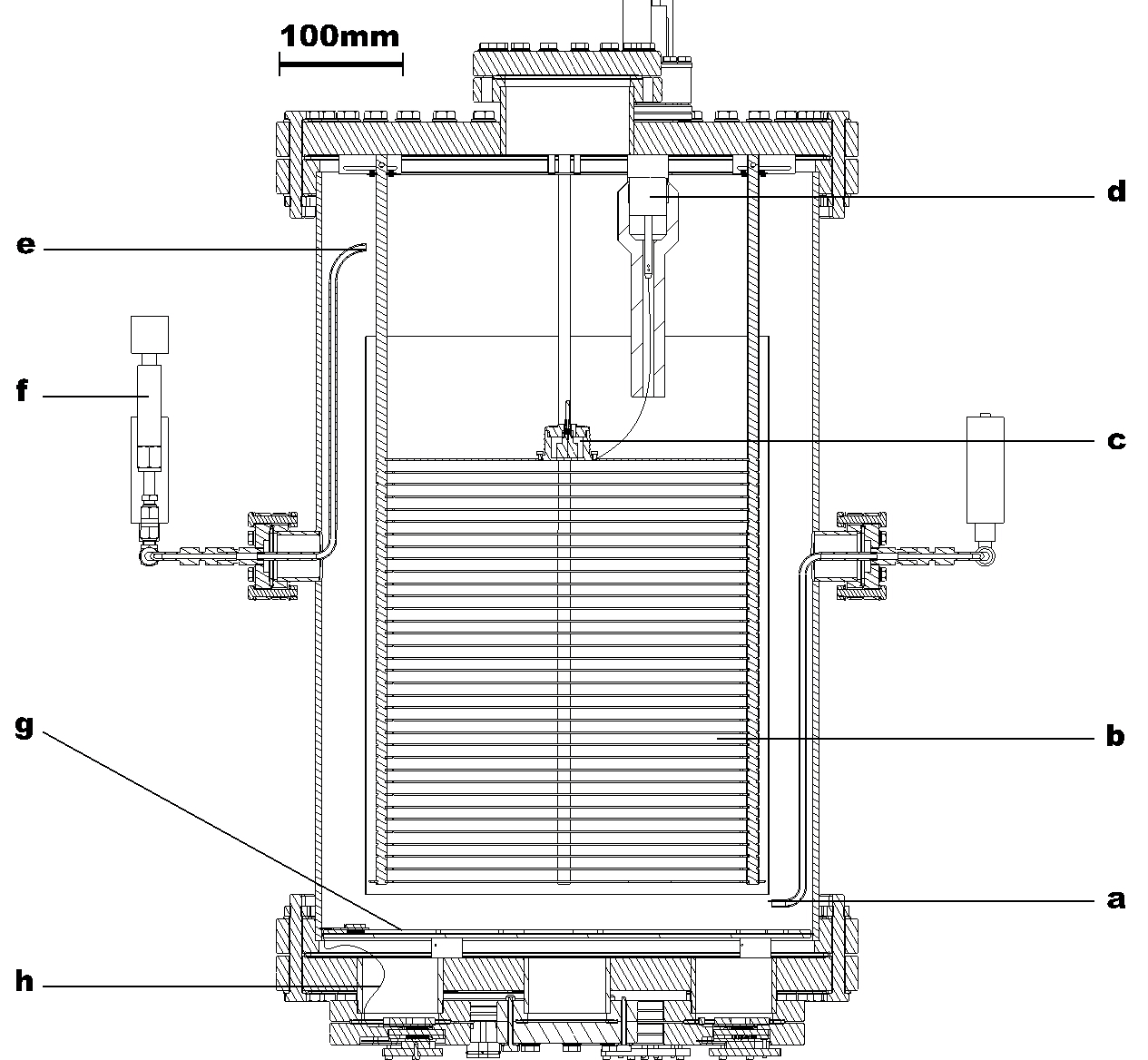}
\caption{Picture (left) and design (right) of the TREX-$\beta\beta$-1 (NEXT-MM) system, indicating the gas inlet (a), field cage (b), internal calibration source (c) high voltage feedthrough (d) gas outlet (e) pressure gauge (f) Micromegas read out plane (g) and flat cable to extract the signals (h). Full description can be found in \protect\cite{Alvarez2014a,Alvarez2014,GonzalezDiaz20158}}
\label{fig:setup}
\end{figure}

An important step towards experimental demonstration of the concept of a Micromegas-TPC in Xe+TMA has been achieved with the TREX-$\beta\beta$-1 setup at the University of Zaragoza. Originally developed as part of the prototyping program for the NEXT experiment, it is referred to as NEXT-MM prototype, and described in detail, in \cite{Alvarez2014a,Alvarez2014,GonzalezDiaz20158}. This prototype, shown in Figure \ref{fig:setup}, is able to host more than $\sim$1 kg of Xe at 10 bar in its sensitive volume. The vessel is built with combined high-pressure and ultra-high-vaccuum specifications and is connected to the necessary gas recirculation system including a gas mixer, recirculation pump, appropriate filters and a mass spectrometer. It also features a number of tailor-made feedthroughs (especially the cathode's high voltage one and several high-density multi-channel signal ones at the readout) that have been used to test technical solutions of possible radiopure implementation of the detection concept. We refer to~\cite{Alvarez2014a} for the technical details.

\begin{figure}[b!]
\centering
\includegraphics[height=45mm]{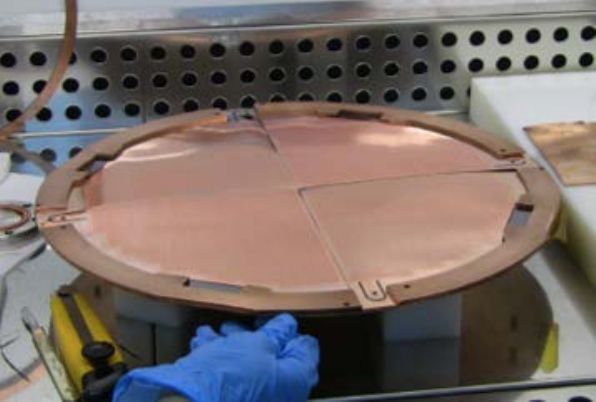}
\caption{Picture of the 4-sector 30-cm diameter microbulk readout referred to in the text.}
\label{fig:mm}
\end{figure}

The sensitive volume has a cylindrical shape, with a drift length of 38~cm and 30~cm of diameter, corresponding to 24 liter, and sufficient to fully contain high energy ($\sim$MeV) electron tracks and thus to characterize the readout at energies closer to \Qbb\ . The system is able to host large-area (up to 30~cm diameter) highly granular readouts. This environment reproduces, in what the readout is concerned, the same operating conditions of a full scale TPC for DBD. Different readouts and gases have been tested, but we review here the main results obtained in this chamber, as part of the NEXT-MM program, involving a microbulk Micromegas of 30 cm diameter in Xe+TMA. This readout is the largest readout surface with this technology operated so far, and was composed of 4 circular sectors, each of them at the time the largest single wafer manufactured with this technology (see Figure \ref{fig:mm}). The readout was pixelated with 8 mm side pixels, making a total of 1152 channels. Each channel was independently read with a DAQ chain based on the AFTER chip, which provides full TPC functionality (i.e. temporal waveform of each of the channels). The microbulk meshes were read as well and used to trigger the pixel DAQ. Apart from the results presented below, this experimental program provided operating experience in a number of key technical aspects, very relevant to the current proposal, like the handling of large number of signal channels and extraction from high pressure vessel, feedthroughs, high voltage in Xe gas, gas recirculation with TMA, gas gain, crosstalk and equalization of channels in microbulk readouts, discharge statistics and related dead time, etc.

Low energy characterization of the chamber was made with the x-ray emission of an $^{241}$Am calibration source, placed inside the chamber and tagging its alpha emission with a small Si-detector, to externally provide the $z$-coordinate of the event exploiting the coincidence between the alpha and the x-ray. High energy electron tracks, of 511~keV and 1275~keV, are generated with a $^{22}$Na calibration source placed outside the chamber. External determination of the $z$-coordinate for the 511~keV is also obtained by tagging the second 511~keV photon in coincidence with a NaI detector. The external determination of $z$ is used to test that no appreciable hint of attachment is present in the data for a wide range of drift fields~\cite{Alvarez2014,GonzalezDiaz20158}.

\begin{figure}[t!]
\centering
\includegraphics[width=0.29\textwidth]{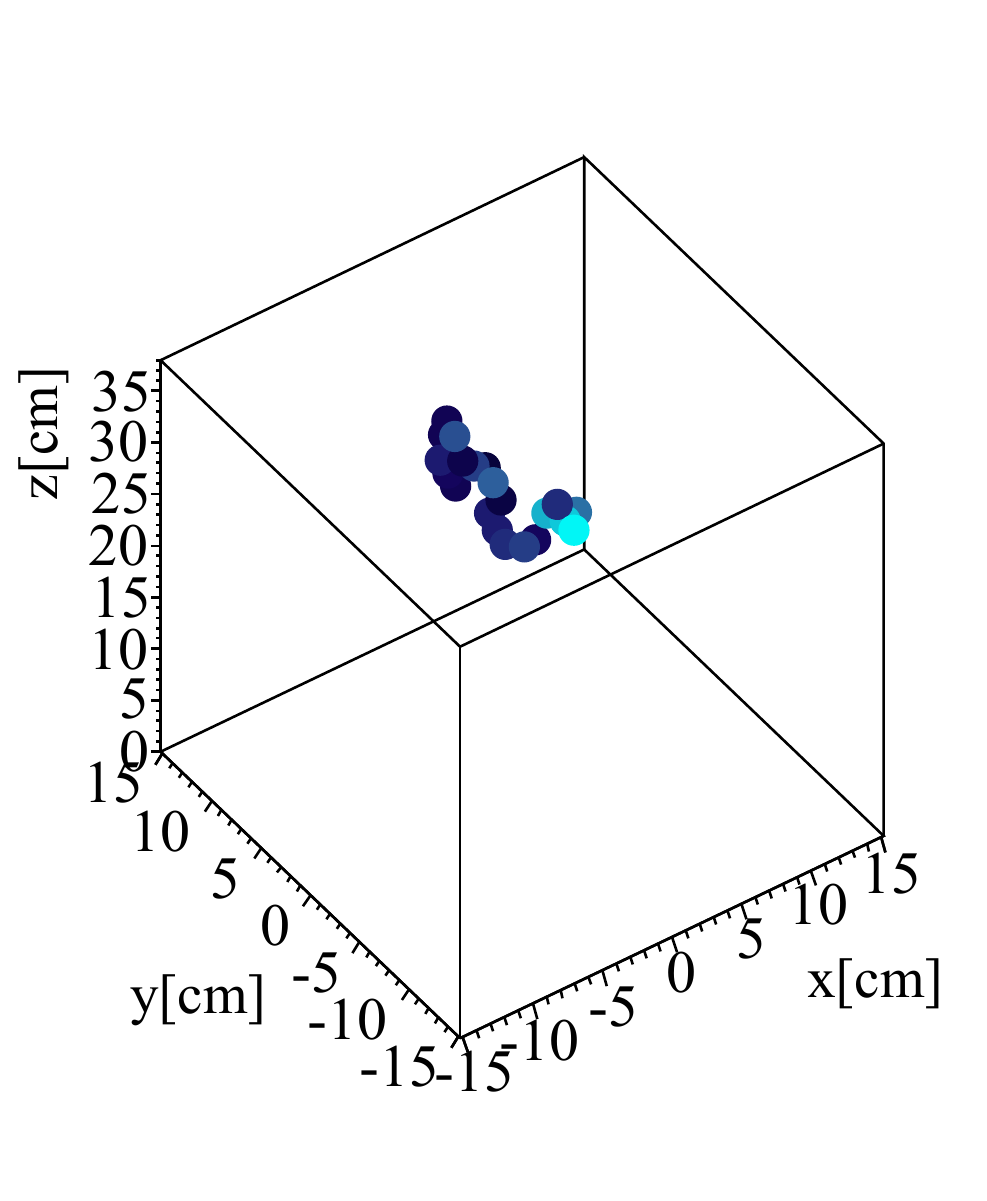}
\includegraphics[width=0.29\textwidth]{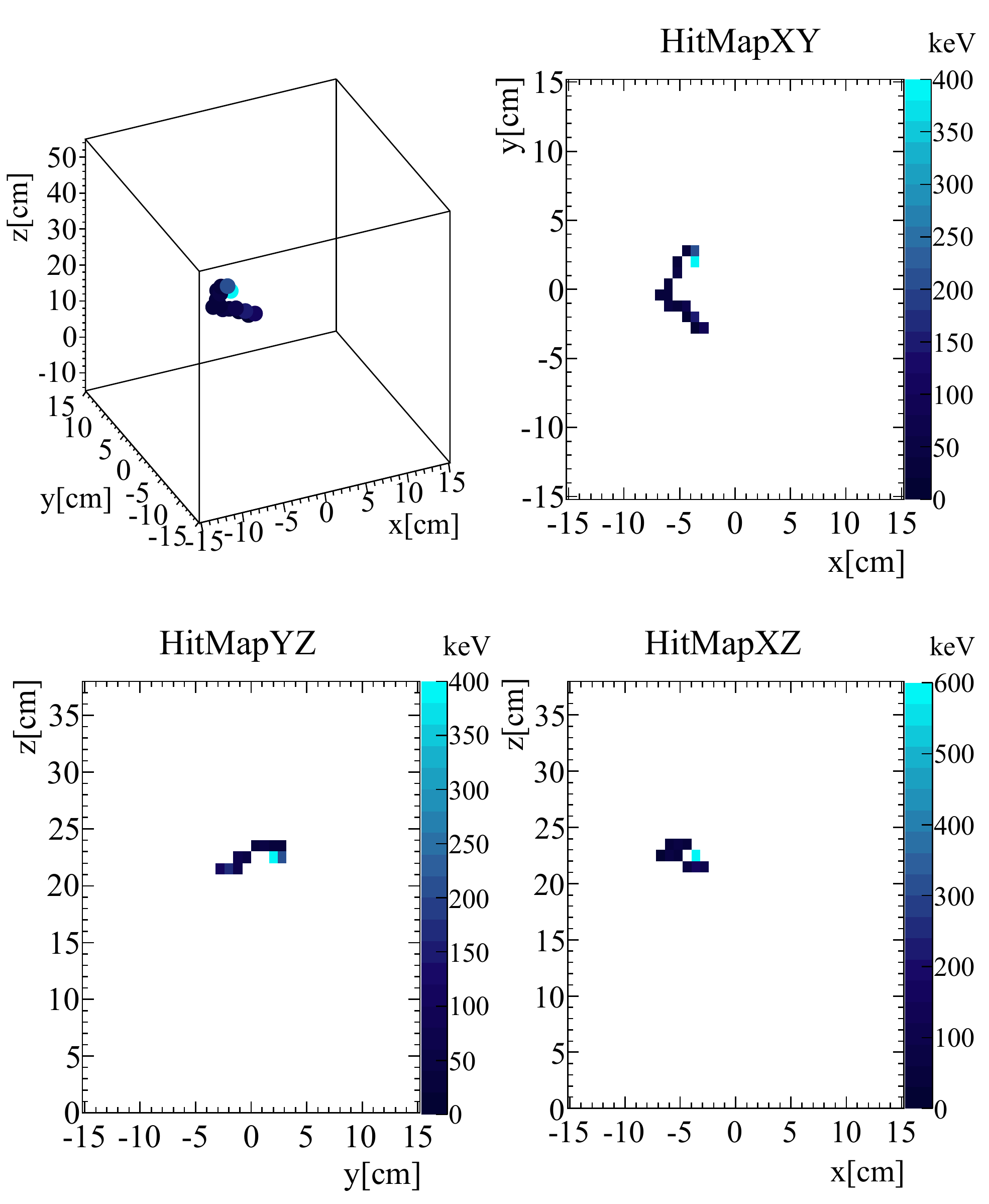}
\includegraphics[width=0.29\textwidth]{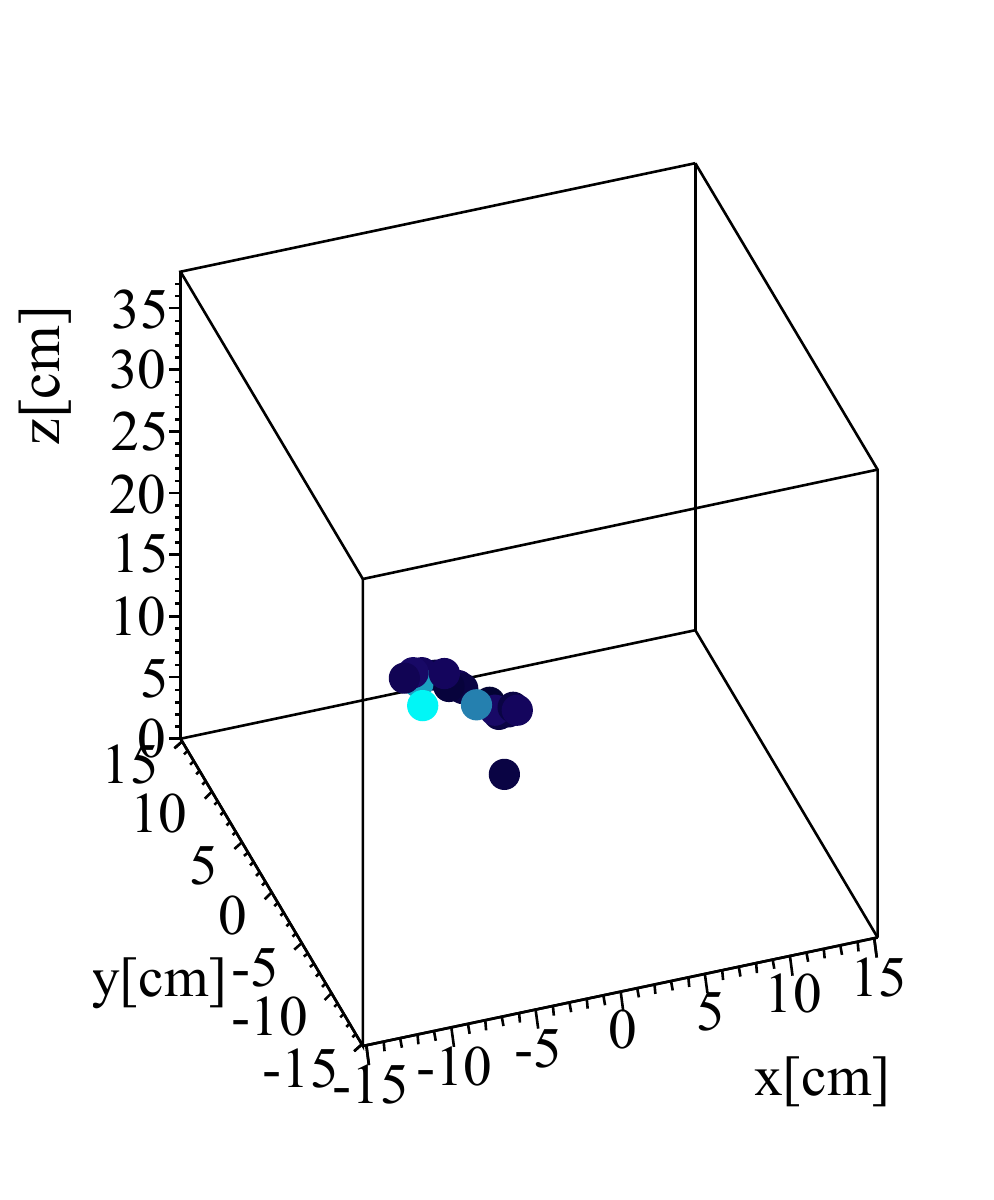}
\includegraphics[width=0.35\textwidth]{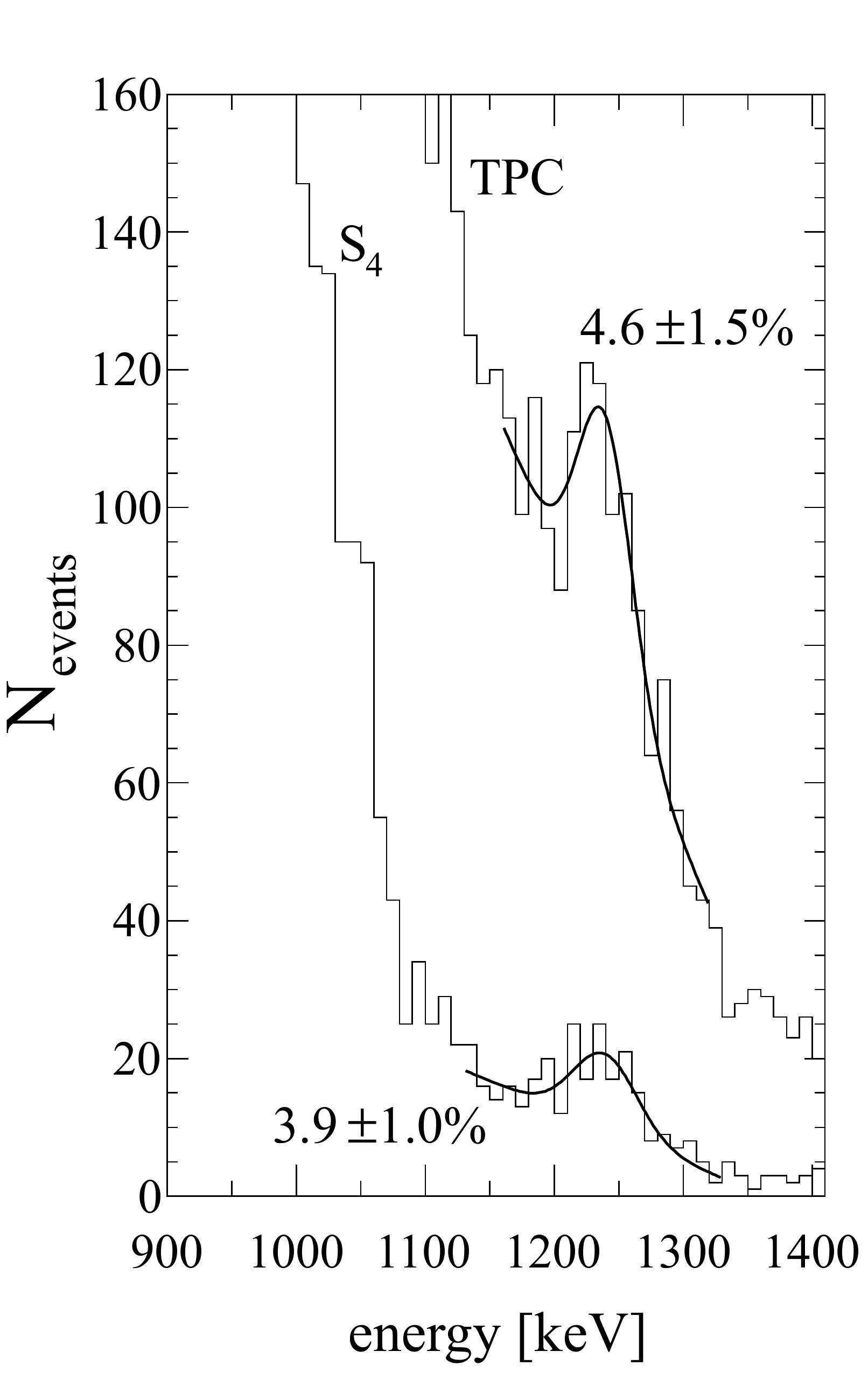}
\includegraphics[width=0.35\textwidth]{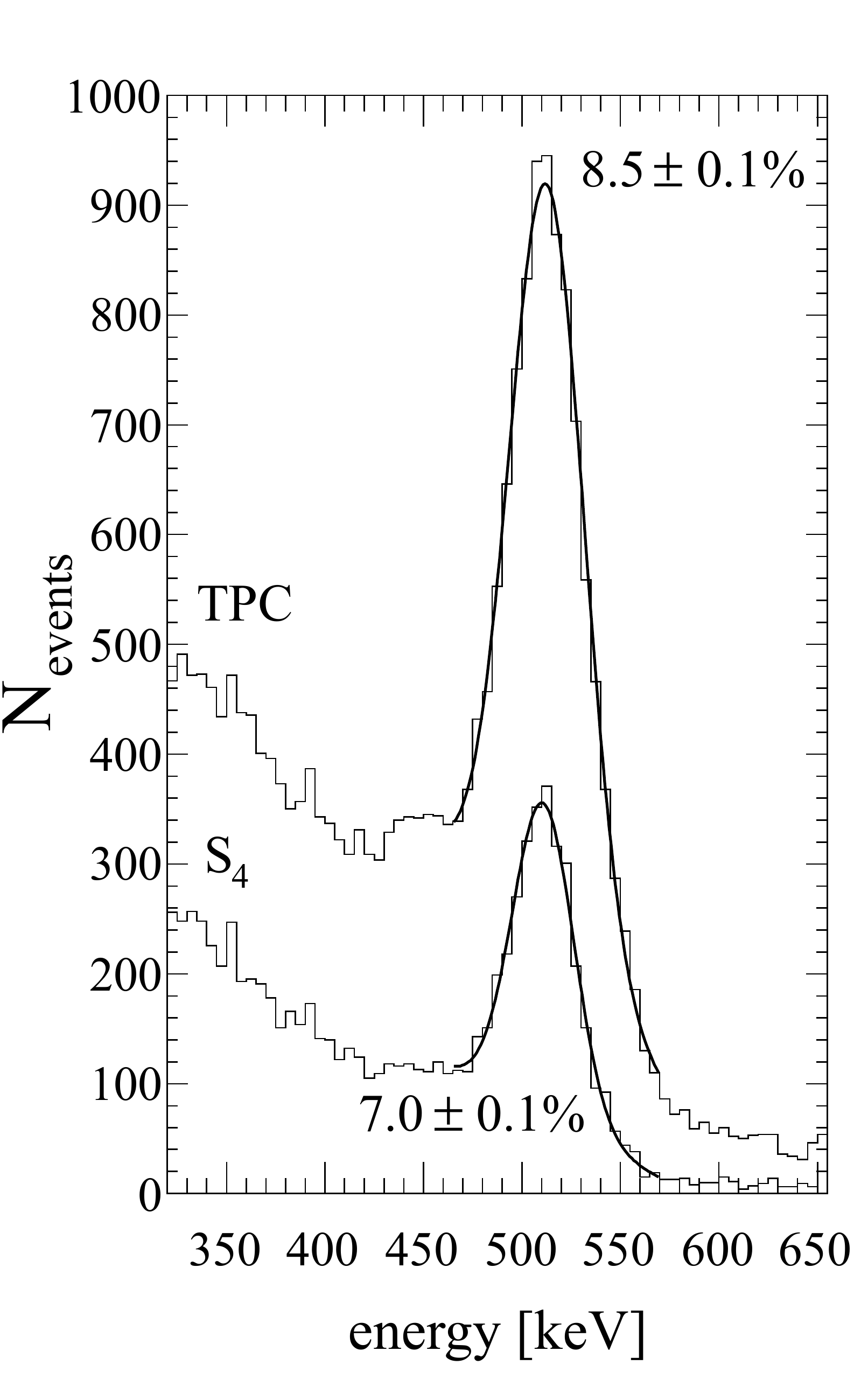}
\caption{Top: Reconstructed 3D tracks from several illustrative 1275~keV events in the T-REX with the NEXT-MM Micromegas, were the end of the track ending in the Bragg peak is well identified (lighter color circles mean higher collected charge).
Bottom: Energy resolution obtained for the two peaks of $^{22}$Na at 511~keV and 1275~keV for the TPC fiducial region (TPC) and for one of the sectors (S4). See~\cite{GonzalezDiaz20158} for more details.}
\label{fig:tracks}
\end{figure}

Figure \ref{fig:tracks} shows some examples of 3D tracks reconstructed from the microbulk signals, corresponding to energies at the 1275~keV peak. It must be noted that these events' energy equals the mean energy of one of the DBD electrons (\Qbb$/2$), and are therefore representative of what (``half'') the DBD signal events would look like. The event topologies show that accurate reconstruction is possible, even if the pixel size is here the limiting factor too (again, a pixel size of 8$\times$8 mm was fixed at an early stage of the project, in view of operation in pure Xe). Despite this limitation, visual inspection of the sub-sample of events containing a single $\sim$1275~keV track showed that the blob is clearly distinguishable for all of the events of the sample~\cite{GonzalezDiaz20158}. A particular analysis using the higher effective granularity in the $z$-direction (determined by the electronics sampling and shaping times, and amounting to about 1.2 mm) confirms the enhanced identification capabilities of the blob features brought by the low diffusion of Xe+TMA, and constitutes a preliminary qualitative confirmation of the arguments exposed in the previous section. Efforts to quantitatively assess them in the context of recognition algorithms like the one described before are ongoing. New experimental campaigns with more adequate pixel size are also planned for the near future.

The energy of the events is extracted by summing the charge from every active pixel in the event after a careful low-level analysis including pixel-to-pixel calibration, control of non-functional pixels, temporal drift of gain, etc. Energy spectra for high statistics $^{22}$Na runs are built, and energy resolutions down to $\sim$7\% and $\sim$4\%~FWHM are obtained for the peaks of 511~keV and 1275~keV respectively (see Figure~\ref{fig:tracks}). We consider this analysis a conservative demonstration of the energy resolutions achievable by pixelated Micromegas readouts for high-energy extended tracks at 10~bar, pointing to values of the order of $\sim$3\% for \Qbb\ . A number of instrumental limitations have been identified as the reasons why this value is still a few times higher than the ones obtained in small readouts and low energies, as presented in section \ref{sec:micromegas}. They include, possibly, a reduced signal-to-noise ratio in the larger setup with respect to the smaller one, the presence of some crosstalk in the readout, and some degree of undersampling in the temporal waveforms. Work to overcome one or more of these issues should help improve the energy resolutions possibly down to levels close to 1\% FWHM. Nevertheless, the values so far obtained are the best for charge readout in Xe (e.g. improving those of \cite{Luscher:1998sd} in about a factor of 2) and are already very competitive for next generation DBD experiments.

\begin{figure}[t!]
\centering
\includegraphics[width=0.5\textwidth]{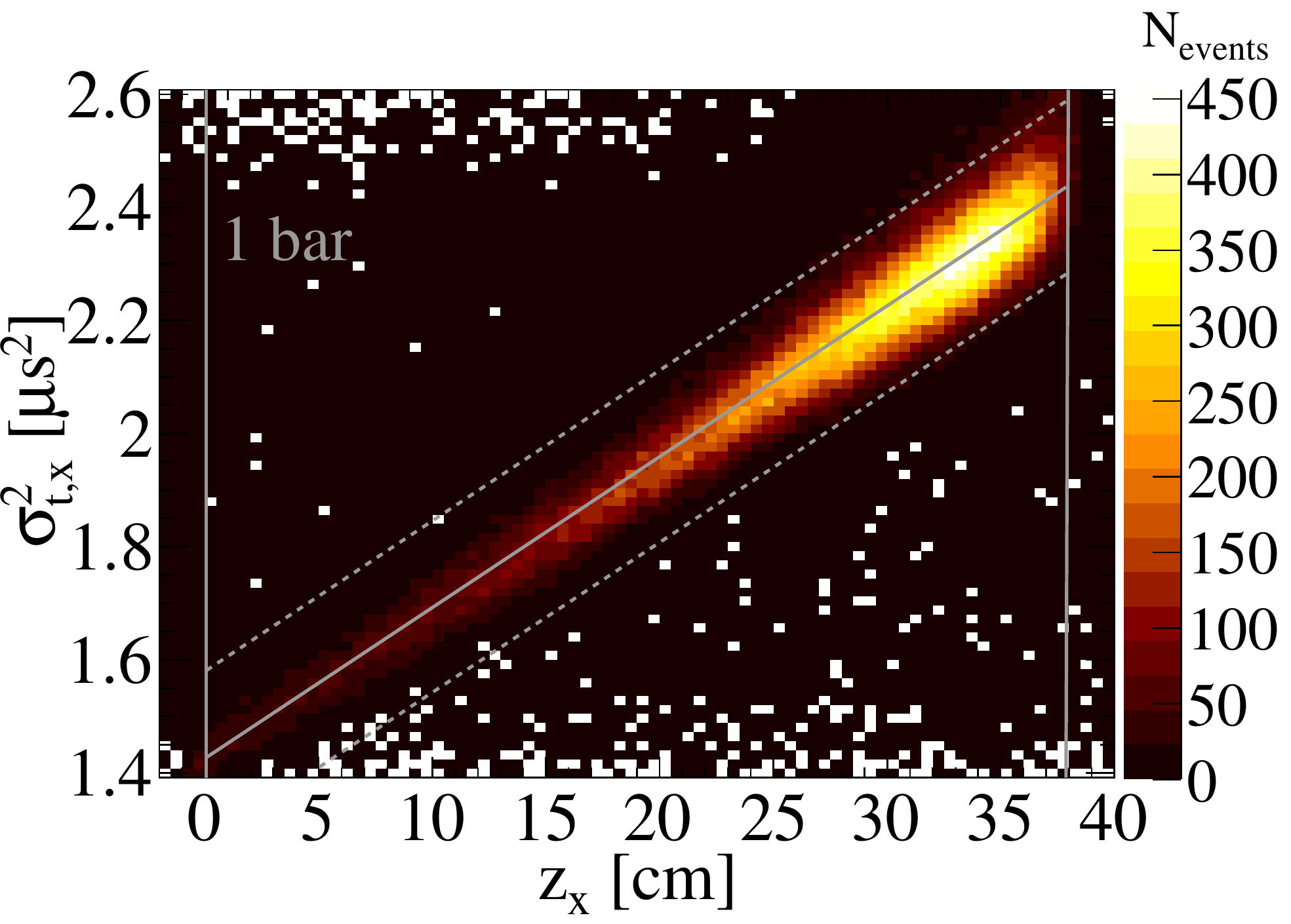}
\caption{Correlation of electron diffusion (width of pulse) versus $z$-coordinate experimentally obtained in low energy x-rays in 1 bar Xe+TMA~\protect\cite{GonzalezDiaz20158}}
\label{fig:diff_vs_z}
\end{figure}

To finalize this section, it is worth to note that these data confirm the properties of Xe+TMA presented in section~\ref{sec:micromegas}, both regarding transversal and longitudinal diffusion coefficients, and drift velocities, which in turn are in reasonable agreement with Magboltz calculations~(see Figure~\ref{fig:diff}). Of particular interest is the relatively sharp correlation between diffusion and $z$-coordinate demonstrated with x-rays at low pressure (1 bar) shown in Figure \ref{fig:diff_vs_z}. This correlation opens the way to extract some approximate absolute $z$-coordinate information from the event topology, in the line of what has been discussed in the previous section, and compensating, at least in part, the absence of $t_0$. This issue is commented further on in section~\ref{sec:discussion}. Similar data at higher pressure are not available due to the difficulty to illuminate all the TPC volume with low-energy x-rays, but in principle, there is no fundamental reason why the concept should not equally work, once the known pressure-scaled drift and diffusion properties are taken into account.


\section{Towards scaling-up}
\label{sec:scaling}

Although very large areas of Micromegas readouts can be built (as mentioned in section~\ref{sec:micromegas}, 1200 m$^2$ are being built~\cite{Losel:2015una} in the context of the ATLAS upgrade), the fabrication technique invoked in these large-area projects is usually the ``bulk'' Micromegas technique~\cite{Giomataris:2004aa}, not yielding -in general- radiopure readouts. The possibility of building radiopure bulk Micromegas is being explored (measurements \#13 and \#19 of Table~\ref{tab:radiopurity} suggest its feasibility). However, in view of the results presented in previous sections, the feasibility of  any large scale project aiming at applying the detection concept here proposed, relies on the possibility of building large areas of \emph{microbulk} Micromegas, keeping the salient features explored in the previous sections, not only radiopurity, but signal quality, energy resolution, granularity, etc. Although a priori there is no fundamental reason why large area microbulk Micromegas cannot be built, some limitations stem from practical issues (e.g. size of available equipment). Although these limitations could be surmounted by some R\&D (as indeed is being done), noise considerations would probably also set a limit to the size of a single microbulk unit that can be realistically used in a large DBD experiment, due to electronic noise considerations. At the moment the largest microbulk Micromegas operated is the 30~cm diameter one built for the present project, presented in section~\ref{sec:etracks}, and composed by 4 independent microbulk sectors. The definition of a realistic (and even conservative) scaling-up strategy to equip a large area, of the order of a few meters, with microbulk Micromegas, has been one of the goals of T-REX.

\begin{figure}[t!]
\centering
\includegraphics[height=65mm, trim={10cm 5cm 0 0}, clip]{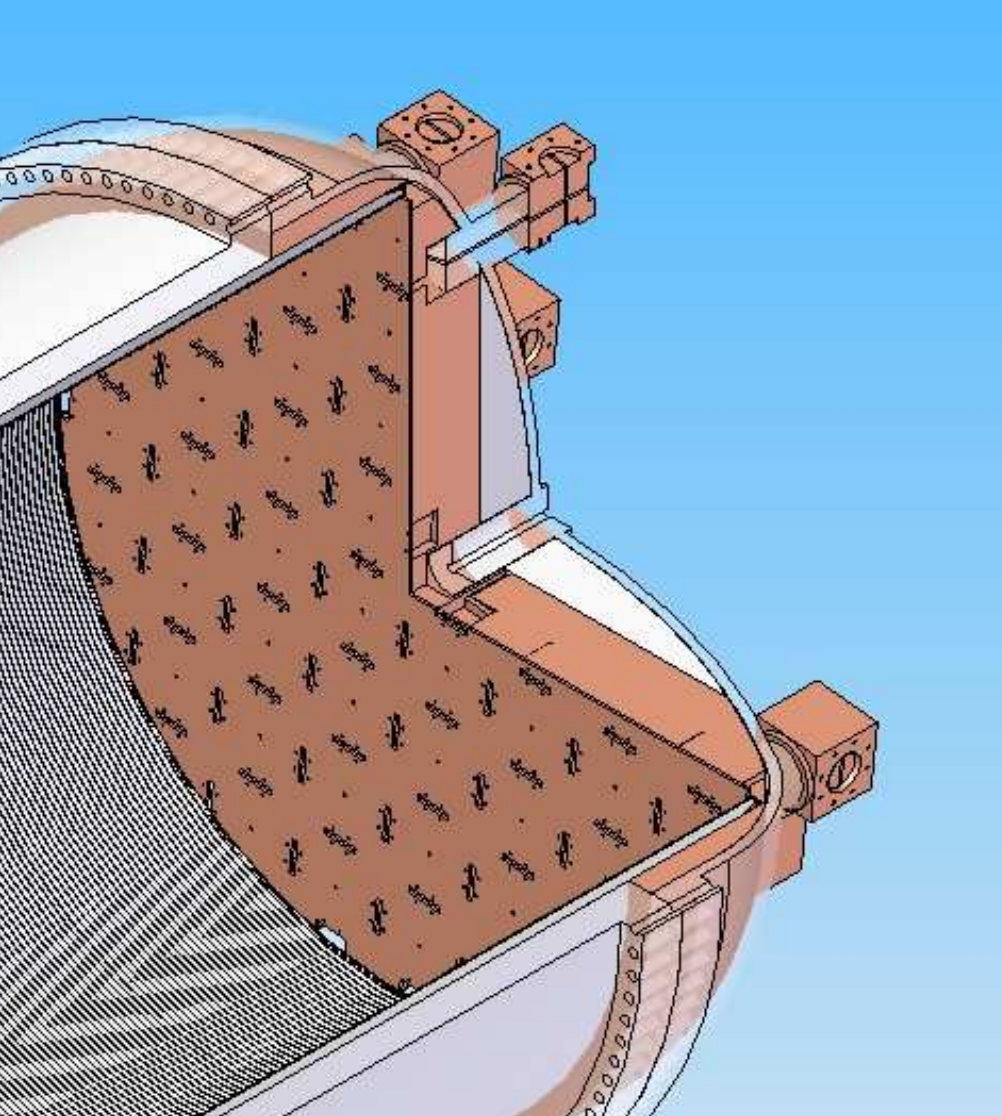}
\includegraphics[height=65mm]{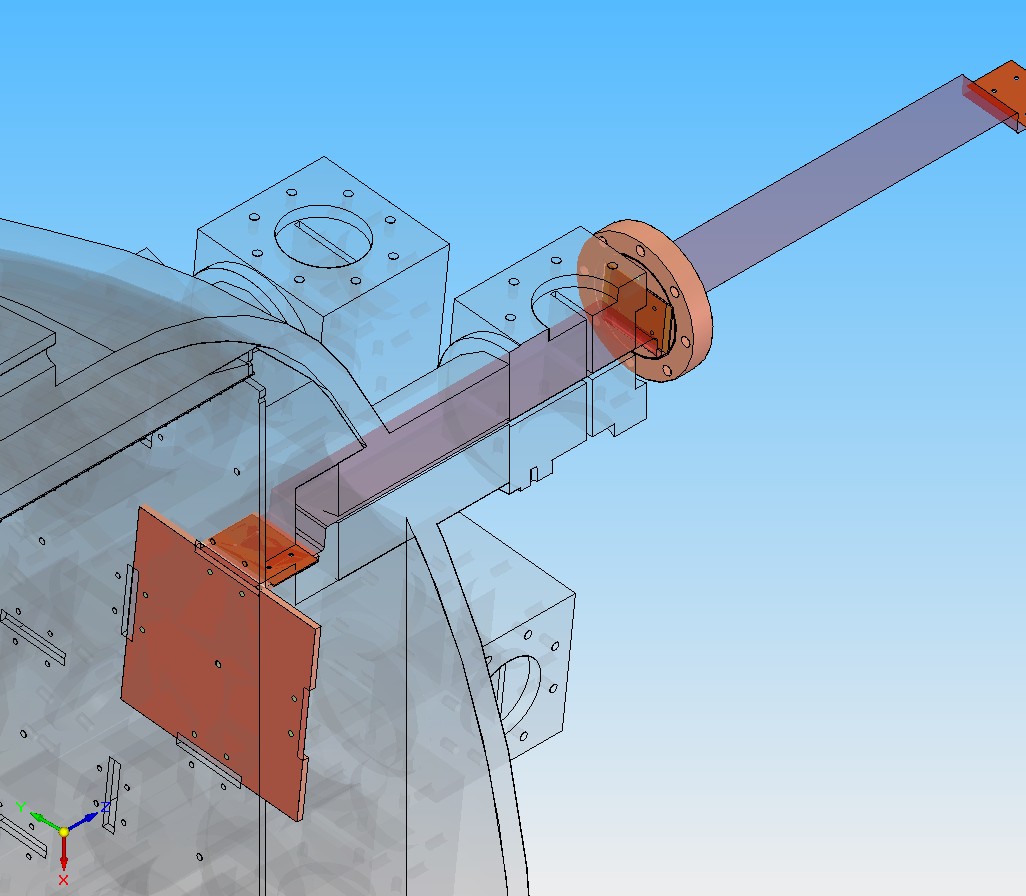}
\caption{Design of a possible mechanical implementation of the SR2M concept (see text). On the left, the support plate of the tessellation of many identical single modules as the one shown on the right. The signal extraction is done extending outwards the microbulk kapton-copper foil, keeping any connector or soldering far from the sensitive volume, possibly, as shown in this sketch, out of the detector vessel.}
\label{fig:SR2M}
\end{figure}

\begin{figure}[t!]
\centering
\includegraphics[height=37mm]{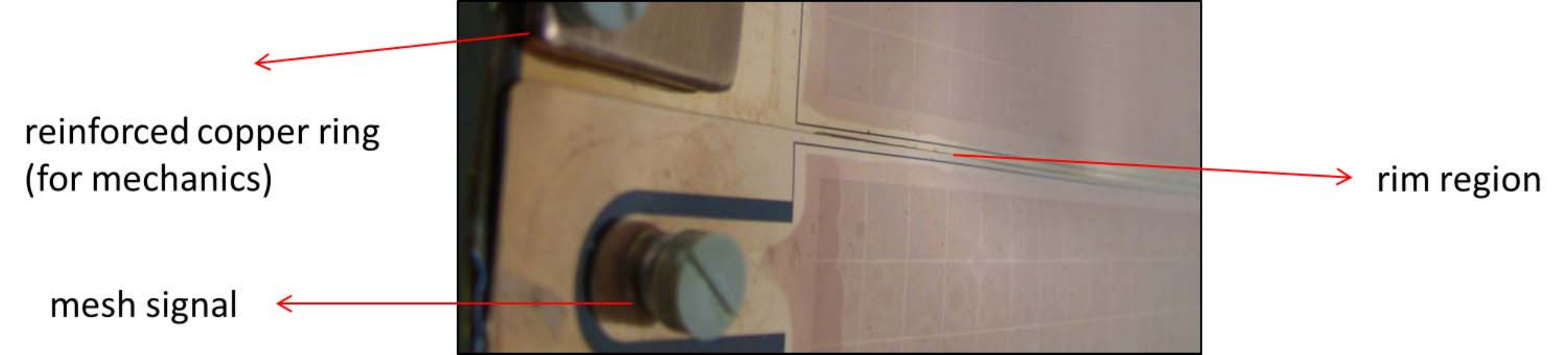}
\caption{Picture of the rim electrode as tested in the NEXT-MM 4-sector microbulk readout \cite{Alvarez:2013kqa}.}
\label{fig:rim}
\end{figure}

The concept put forward here is based on the tessellation of many relatively small identical modules to cover a large area. Each of the unit modules and the associated mechanics, routing out of the chamber and electronics would be replicated, facilitating the engineering of the full size detector. In this way, the size and design complexity of the single unit, dubbed Scalable Radiopure Readout Module (SR2M), is similar to the ones already experimentally tested here, and therefore well within current state-of-the-art. The main elements of the scaling-up strategy are:

\begin{itemize}
  \item The size of each SR2M microbulk can be comparable (e.g. 20$\times$20~cm$^2$) to the areas already successfully fabricated and operated within T-REX.
  \item The routing of the signal channels of each SR2M unit are extracted via a flexible cable that is the continuation of the very same kapton-copper microbulk foil (see figure~\ref{fig:SR2M}), avoiding the need of connectors, soldering or any other material close to the sensitive volume. The foil/cable is bent outwards as soon as it goes out of the Micromegas active area, as shown on the right of Figure~\ref{fig:SR2M}, without conflict with the tessellation, and bringing the signals far enough from the readout so that any additional interfaces or feedthroughs could --if needed-- be shielded from the active volume.
  \item The SR2M microbulk will be glued on to a radiopure copper support structure, to give mechanical strength to the readout, and provide good electronics grounding/screening, while keeping the highest radiopurity constraints.
  \item The SR2M has a margin of $\sim$1~mm from the active area to the edge of the module, due to fabrication constraints, which constitutes a dead area for detection. Limited assembling tolerance between modules, and the bending of the kapton foil to extract the signals also contribute to this dead area. Its presence may affect the energy resolution for DBD electron tracks traversing more than one module. To solve the problem of these border effects, a thin electrode surrounding the active area of each module is engraved (the ``rim'' electrode) and is independently powered at a higher voltage than the mesh voltage. This rim gently pushes the drift lines away from the dead areas and drives them toward the active areas. A first version of the rim concept has already been tested in the 4-sector configuration described in section~\ref{sec:etracks}, see Figure~\ref{fig:rim}.
\end{itemize}

Figure~\ref{fig:SR2M} illustrates the SR2M concept in a generic way. Further technical development of the concept must be carried out within the scope of a specific experimental implementation. This will be done in the near future within the recently proposed PandaX-III experiment. In any case, the concept status shown by T-REX presents sufficient evidence of the feasibility of a few-m$^2$ area microbulk readout. The idea relies mostly on known and tested solutions, and justifies the statement advanced in the previous section, regarding the simplicity of this detection configuration in terms of number of materials close to the sensitive volume.



%
%

\section{Discussion and outlook}
\label{sec:discussion}

Throughout the previous sections, evidence has been presented supporting the feasibility and competitiveness of a microbulk Micromegas as the charge readout of a HPXe TPC for the search of the $^{136}$Xe DBD decay. Aspects like radiopurity, background rejection by topological recognition, energy resolution and scalability have been explored. The use of TMA as an additive to the Xe is an important ingredient of the concept (with definite impact in the energy resolution, stability of operation and the topological quality) but it comes at the price of quenching the scintillation of the Xe, and, in principle, preventing the measurement of the $t_0$ of the event. We want to discuss in the following what the real impact of this lacking is, as well as more generally the advantages of the detection concept here advocated.

In principle, $t_0$-information allows fiducialization in the $z$-direction (in the $x$ and $y$ direction it is provided by the readout pixelization). This is considered desirable for two aspects:  1) to get an additional handle to reject background; and 2) to be able to correct for attachment on an event-by-event basis, in the case it is present. We argue in the following that, under the light of the work presented in the previous sections, these reasons are of little weight in the experiment's final figure of merit.

While event fiducialization is very important in dark matter and DBD \textit{liquid} Xe experiments, it turns out to be of very little impact in gas TPCs. This is due to the absence of self-shielding in gas. Indeed, as quantified in the studies presented in section~\ref{sec:topology}, the discriminating effect of the fiducial cut on background is practically negligible, with the \textit{only} exception of very specific background populations: those coming from surface contaminations of the $\beta$-emitter $^{214}$Bi isotope (and  \emph{only} those present in the cathode and anode of the TPC). The final effect on the background level from the lack of $t_0$ will in general depend on the importance of this particular background population in the overall background budget of the experiment, but will generally be mild. Conversely, one can argue that absence of $t_0$ is then acceptable if these particular populations are well under control. The situation will certainly be much better than that, because a very rough $z$-information coming from e.g. diffusion (see discussion later on), will already be sufficient to reject this background.

Regarding the issue of attachment, current experimental data have explored drift distances up to 38~cm in 10~bar Xe+TMA mixtures, with no evidence of attachment effects. This distance is lower but already representative of the drift distances involved in a 100~kg scale experiment. Therefore it is safe to conclude that the lack of attachment-correction capabilities due to absence of $t_0$ will not have a relevant impact on the experiment's performance (e.g. energy resolution).

On the other hand, the absence of photosensors in the detector configuration brings important advantages, e.g. reducing the detector complexity and its radioactivity budget. It is known that, despite the large efforts in reducing their radiopurity, photosensors (e.g. PMTs) are expected to be the main source of background in next generation rare event searches\cite{Baudis:2013qla}. The detector configuration here proposed, avoiding them completely, will give access to lower background levels that surely compensate any negative factor from the above drawbacks in the final experiment figure of merit.

It must be noted, that even in the absence of a scintillation signal, $t_0$ information on an event-by-event basis can be extracted from the diffusion properties of the event. Figure~\ref{fig:diff_vs_z} shows the correlation found experimentally between diffusion (width of electron cloud) and $z$-coordinate in low energy events at 1 bar. These data suggest that diffusion can in principle be used to get a rough $z$-coordinate information. This $z$ determination has an uncertainty that comes from the spread in the electron cloud's width resulting from the electron cloud statistics, i.e. the energy of the event. The correlation should a priori become sharper than in Figure~\ref{fig:diff_vs_z} for the higher energy DBD events. However, the capability to efficiently extract diffusion information out of the convoluted high-energy track topology remains to be proven technically. The study of some topological observables showing a clear link with diffusion, like the blob density studied in section \ref{sec:topology} and Figure~\ref{fig:TopRejection} is to be considered a first step in this direction.

To conclude, the concept of a pure-charge Micromegas-read TPC is an elegant and powerful detection concept for the search of $^{136}$Xe DBD. The concept allows for a simple implementation involving a very reduced material budget. Critical (i.e. unavoidable) materials are basically reduced to the materials composing the microbulk readouts (mainly kapton and copper), together with the material composing the detector vessel and inner field-cage structure, conceivably reducible to e.g. teflon and copper.  No other constraints regarding needed materials or critical components exist that could limit the detector radiopurity. Despite the conceptual complexity of a TPC, with several thousands of readout channels, the current implementation avoids connectors, soldering or electronics components close to the sensitive volume. Current state-of-the-art of the topological recognition algorithms, together with the maximum radioactivity levels expected (based on upper limits) from the aforementioned unavoidable materials represent a level of background of $\sim$10$^{-4}$ c~keV$^{-1}$~kg$^{-1}$~yr$^{-1}$, that corresponds to less than 1(5) count(s) for an exposure of 100(500) kg-year in an energy RoI of 3\% around \Qbb\, and therefore very close to the ultimate sensitivity that an experiment at the $\sim$100~kg scale can provide.

Regarding the subsequent jump to ton-scale DBD experiments, it is acknowledged that much lower background levels, down to $\sim$0.1 counts~ton$^{-1}$~year$^{-1}$ in the RoI, need to be assured. This requirement poses very serious radiopurity challenges to all the candidate detection techniques. Although more work is needed to assess wether this goal is at reach of a HPXe Micromegas-TPC, the results here presented have not identified intrinsic limitations for additional background reduction. Indeed, the needed factor may be at reach by means of one or more of the following issues, already commented along the previous sections: 1) the current background estimation is based entirely on upper limits to the radioactivity of key materials and it may be very well possible that more sensitive measurements reveal that their intrinsic radioactivity is substantially lower; 2) topological cuts can certainly be further refined with respect to the rejection factors presented in this paper, maybe even an extra factor 10 (see section~\ref{sec:topology}); 3) energy resolution could eventually be improved by a factor 3 (see section~\ref{sec:micromegas}); and 4) an extra factor of $\sim$2 is expected simply by reduced surface-to-volume ratio in a 10 times scaled-up detector.

\section{Conclusions}
\label{sec:conclusion}
We have demonstrated that a microbulk Micromegas Xenon TPC is a very competitive modern incarnation of the HPXe TPC concept pioneered by the Gotthard group 20 years ago. Micromegas of the microbulk type can be built and implemented with extremely low levels of radiopurity, below 0.1$\mu$Bq/cm$^2$, which correspond to a negligible component in the radioactive budget of current DBD experiments. They show very good performance in Xe+TMA mixtures at 10 bar, with good figures in terms of gain, energy and spatial resolution.
Operation in realistic experimental conditions (10 bar Xe+TMA and large enough detector size: 30 cm diameter readout area, 1200 readout channels, 38 cm drift)
has proven energy resolution of about 3\% FWHM at the $Q_{\beta\beta}$ value for high energy long electron tracks, a value that improves that of the  Gotthard TPC by a factor $\sim$2. This value is however limited by practical and not fundamental reasons. Energy resolutions extrapolating to 1\% FWHM @ $Q_{\beta\beta}$ at 10 bar Xe+TMA have been obtained in a smaller setup with low energy sources. In addition, Xe+TMA shows transversal diffusion corresponding to about $\sim$1 mm-$\sigma$ for 1~m drift, a remarkable value that enhances the quality of the event topology. Microbulk Micromegas can easily be patterned with sufficient granularity to match this low diffusion. The discrimination from background by inspecting the DBD topological information (a straggling line with two blobs) is expected to effectively reject background in the RoI by, at least, about an extra order of magnitude with respect detectors with only monosite/multisite discrimination capabilities (like liquid Xe), while there are hints that this rejection can be higher.

The background level anticipated from the above results, for a 100~kg scale experiment, is of less than 1 background count in the RoI for an exposure of 100 kg-year. Technical solutions to implement this readout concept at scales of the few m$^2$ have been developed, keeping the state-of-the-art performance obtained in the small scale prototypes. Therefore we conclude that a large ($\sim$100~kg) scale TPC equipped with microbulk Micromegas operating in 10 bar Xe-(1\%)TMA will be a very competitive DBD experiment. These concepts will be realized by the PandaX-III collaboration, that has recently started the project to build one such 200~kg detector. Clear margin of improvement exists in radiopurity limits, topological discrimination and energy resolution Although additional future work is needed to define the extent of these additional improvements, enough motivation exists to consider this detection technique for the jump to the ton-scale experiments.

\acknowledgments

The authors would like to warmly acknowledge the many collaborators that have contributed to the T-REX results obtained over the last years, from the GIFNA group, from CEA/Saclay, CERN and LSC, as well as form the CAST, IAXO, NEXT, RD-51 and PandaX-III collaborations for many motivating discussions. We want to specially thank D. González-Díaz. Our special thanks go also to Rui de Oliveira workshop at CERN for fabricating for us the microbulk readouts, and CEA/Saclay colleagues for their always wise advice. We thank the PandaX-III collaboration for the opportunity to apply at large scale the concepts explored and tested within T-REX in the near future. We acknowledge the use of the Servicio General de Apoyo a la Investigación-SAI, Universidad de Zaragoza. All this research has been mainly supported by the ERC Starting Grant T-REX ref. ERC-2009-StG-240054 of the IDEAS program of the 7th EU Framework Program. Support is acknowledged also from the Spanish Ministry of Economy and Competitiveness under grants FPA2008-03456, CSD2008-0037, FPA2011-24058, and FPA2013-41085-P, and the University of Zaragoza under grant JIUZ-2014-CIE-02. F.I. acknowledges the support from the \emph{Juan de la Cierva} program and T.D from the \emph{Ram\'on y Cajal} program of the MICINN.

\bibliographystyle{jhep}

\bibliography{trexbb}
\end{document}